\documentclass[prl,twocolumn,nofootinbib,aps,tightenlines,preprintnumbers,notitlepage,longbibliography,superscriptaddress]{revtex4-1}
\usepackage{graphicx}
\usepackage{amsmath}
\usepackage{amsfonts}
\usepackage[colorlinks=true,citecolor=blue,urlcolor=cyan,linkcolor=black]{hyperref}

\newcommand{\nhat}{\hat{ \mathbf{n}}}

\usepackage{amsmath}
\usepackage{amsfonts}

\usepackage{tabularx}
\newcolumntype{C}{>{\centering\arraybackslash}X}
\newcolumntype{R}{>{\raggedleft\arraybackslash}X}

\usepackage[usenames, dvipsnames]{color}
\usepackage[normalem]{ulem}
\usepackage{xcolor}

\newcommand{\dd}{{\rm d}}

\def\nhat{\hat{\mathbf{n}}}
\def\khat{\hat{\mathbf{k}}}

\def\khat{\hat{\mathbf{k}}}

\newcommand{\be}{\begin{eqnarray}}

\newcommand{\ee}{\end{eqnarray}}

\definecolor{colorA}{HTML}{1E90FF}
\definecolor{colorB}{HTML}{228B22}
\definecolor{colorC}{HTML}{FF7F00}
\definecolor{colorD}{HTML}{4B0082}
\definecolor{colorE}{HTML}{B22222}

\definecolor{lgreen}{HTML}{32CD32}
\definecolor{lgray}{HTML}{D3D3D3}
\definecolor{dblue}{HTML}{1E90FF}
\definecolor{dblue}{HTML}{1E90FF}
\definecolor{orange}{HTML}{FF4500}
\definecolor{indigo}{HTML}{4B0082}

\definecolor{teal}{HTML}{008080}
\definecolor{firebrick}{HTML}{B22222}
\definecolor{salmon}{HTML}{FA8072}
\definecolor{darkgreen}{HTML}{006400}

\usepackage{multirow}
\usepackage{todonotes}

\newcommand{\perimeter}{Perimeter Institute for Theoretical Physics, 31 Caroline St N, Waterloo, ON N2L 2Y5, Canada}

\newcommand{\york}{Department of Physics and Astronomy, York University, Toronto, ON M3J 1P3, Canada}

\graphicspath{ {plots/} }

\begin{document}

\title{Constraints on cosmology beyond $\Lambda$CDM with kinetic Sunyaev Zel'dovich velocity reconstruction}

\author{Jordan Krywonos}
\affiliation{\perimeter}
\affiliation{\york}

\author{Selim~C.~Hotinli}
\affiliation{\perimeter}

\author{Matthew~C.~Johnson}
\affiliation{\perimeter}
\affiliation{\york}

\begin{abstract}

Kinetic Sunyaev Zel'dovich velocity reconstruction uses the statistically anisotropic cross-correlation between cosmic microwave background (CMB) temperature anisotropies and a galaxy survey to reconstruct the remotely observed CMB dipole. Using a reconstruction based on data from \textit{Planck} and unWISE, we rule out non-linear Gpc-scale voids, provide the tightest constraint on the intrinsic dipole ($<14 \ {\rm km/s}$ at $68\%$ confidence), rule out matter-radiation isocurvature as an explanation of discrepancies between the measured CMB and galaxy number count dipoles, and constrain the amplitude of local-type primordial non-Gaussianity ($-220\lesssim f_{\rm NL}\lesssim 136$ at $68\%$ confidence) and compensated isocurvature ($-147\lesssim A_{\rm CIP} \lesssim 281$ at $68\%$ confidence). This representative set of constraints on beyond-$\Lambda$CDM scenarios demonstrates the breadth of fundamental science possible with measurements of secondary CMB anisotropies such as the kinetic Sunyaev Zel'dovich effect.

\end{abstract}

\maketitle

Secondary cosmic microwave background (CMB) anisotropies, sourced by the gravitational and electromagnetic interactions of CMB photons with large scale structure (LSS), provide a means to extract information about the distribution of LSS on the largest observable scales, providing a new window on fundamental physics in the early- and late-Universe. While we have measured the primary CMB near the cosmic variance limit with the \textit{Planck} satellite~\cite{PlanckCollaboration2020}, ground-based CMB experiments such as Atacama Cosmology Telescope (ACT)~\cite{Louis_2017}, South Pole Telescope (SPT)~\cite{Carlstrom_2011} and Simons Observatory (SO)~\cite{Ade2019} are dramatically improving measurements of the secondary CMB.

On small ($\sim$ arcmin) angular scales, the kinetic Sunyaev Zel'dovich (kSZ) effect~\cite{Sunyaev1980}, Thomson scattering of CMB photons from free-electrons in bulk motion, is the dominant blackbody temperature anisotropy. The kSZ effect has been detected at high significance using a variety of techniques~\cite{Hand2012,Hill:2016dta,DES:2016umt,ACTPol:2015teu,Bernardis2017,Tanimura:2020une,Kusiak:2021hai,AtacamaCosmologyTelescope:2020wtv,Chen2022,ACT:2024vsj}, and measurements are expected to improve dramatically in the near future~\cite{Ade2019}. Cosmological information can be efficiently extracted from the kSZ effect using a quadratic estimator for the remote dipole field~\cite{Zhang:2015uta,Terrana2017} -- the locally observed CMB dipole projected along our past light cone (PLC) -- based on the statistically anisotropic cross-correlation of CMB temperature anisotropies and a galaxy redshift survey~\cite{Terrana2017,Deutsch2018a,Smith:2018bpn}. This formalizes the technique of `kSZ tomography'~\cite{Ho:2009iw,Zhang_2010,Shao_2011,Munshi:2015anr}, which we will refer to here as `kSZ velocity reconstruction'. 

Previous work developed complementary methodologies for kSZ velocity reconstruction~\cite{Terrana:2016xvc, Deutsch:2017ybc, Deutsch2018a, Smith:2018bpn, Cayuso2018, Cayuso:2021ljq, Contreras:2022zdz}, and applied analysis pipelines to simulations~\cite{Giri:2020pkk, Cayuso:2021ljq}. The prospects of measuring a wide range of fundamental physics signatures have been assessed in Refs.~\cite{Zhang:2015uta, Munchmeyer:2018eey, Contreras:2019bxy, Hotinli:2019wdp, Cayuso:2019hen, Pan:2019dax, Hotinli:2022jna, AnilKumar:2022flx, Kumar:2022bly, Hotinli:2022jnt, Vanzan:2023cze, Coulton:2023lqe}. Recently, Ref.~\cite{Bloch:2024} applied kSZ velocity reconstruction to data from \textit{Planck} and the unWISE galaxy catalogue to produce a map of the remote dipole field averaged over a redshift range $0.2 \lesssim z \lesssim 1.0$. Although this map is dominated by reconstruction noise, it achieves a sensitivity on large angular scales of $\mathcal{O}(20 \ \mathrm{km/s})$, which is comparable in magnitude to the expected signal within $\Lambda$CDM. In this {\em letter}, we utilize the results of Ref.~\cite{Bloch:2024} to derive the most powerful existing constraints on cosmological models with large-scale inhomogeneities (e.g. voids) and super-horizon matter-radiation isocurvature. We also perform a search for local-type primordial non-Gaussianity and baryon-dark matter isocurvature, providing an important proof-of-principle that kSZ velocity reconstruction will be a powerful probe of these signals with future datasets. A number of technical details are expanded on in the Supplementary Material (SM).

\noindent \textbf{\textit{kSZ velocity reconstruction:}}
The kSZ temperature anisotropies are
\begin{equation}
\Theta_{\rm kSZ}=-\int\dd\chi\,  \dot{\tau}(\nhat\chi) v(\nhat\chi) .
\end{equation}
The differential optical depth is $\dot{\tau}(\nhat\chi) = \sigma_T a(\chi) n_e(\nhat\chi)$ where $\nhat$ is the line-of-sight, $\chi$ is comoving radial distance, $\sigma_T$ is the Thomson cross-section and $n_e(\nhat\chi)$ is the (inhomogeneous) free electron density. For sub-horizon adiabatic modes the remote dipole field $v(\nhat\chi)$ is primarily a Doppler component proportional to the radial peculiar velocity field $\mathbf{v}$, $v(\nhat\chi) \simeq \nhat \cdot \mathbf{v} (\nhat\chi)$; additional contributions from last-scattering are relevant for super-horizon modes. A detailed discussion of signal contributions is presented in the SM.

The remote dipole field is reconstructed from CMB temperature $\Theta(\nhat)$ and projected galaxy density $\delta^g (\nhat)$ using the quadratic estimator~\cite{Bloch:2024} 
\begin{eqnarray}\label{eq:quadestim}
    \hat{v} (\nhat)= - N \left[ \sum_{\ell m} \frac{\Theta_{\ell m}}{C_\ell^{TT,\rm obs}} Y_{\ell m}(\nhat)\right] \left[ \sum_{\ell m} \frac{C_\ell^{\tau g}\delta^g_{\ell m}}{C_\ell^{g g,\rm obs}} Y_{\ell m}(\nhat)\right] 
\end{eqnarray}
\be\label{eq:reonnoise}
1/N=\sum\limits_{\ell'}\frac{2\ell'+1}{4\pi}\frac{(C_{\ell'}^{\tau g})^2}{C_{\ell'}^{g g,\rm obs}C_{\ell'}^{TT,\rm obs}}\,.
\ee
Here, $C_\ell^{TT,\rm obs}$ and $C_\ell^{g g,\rm obs}$ are the observed CMB temperature and galaxy density angular power spectra, respectively. $C_\ell^{\tau g}$ is a model for the galaxy-optical--depth cross power spectrum evaluated at the median redshift of the galaxy survey.
The normalization $N$ is also the estimator variance/reconstruction noise. We ignore the scale dependence of the reconstruction noise, which is an excellent approximation for $\hat{v} (\nhat)$ on large angular scales (see Refs.~\cite{Deutsch:2017ybc,Cayuso:2021ljq,Giri:2020pkk}). 

The estimator mean is
\begin{equation}\label{eq:estim_mean}
    \langle \hat{v}\left(\nhat\right) \rangle = b_v \int d\chi \ W_v(\chi)  \ v(\nhat,\chi) = b_v v\left(\nhat \right) \ ,
\end{equation}
$W_v(\chi)$ is a window function that traces the redshift-dependence of the galaxy-optical depth cross-power spectrum: $W_v \propto C_{\ell=\bar{\ell}}^{\dot{\tau} g}\left(\chi\right)$ where $\bar{\ell} = 2000$ is a fixed  reference scale. Here,  $v(\nhat,\chi)$ is the true underlying remote dipole field, and $b_v$ is the `optical depth bias' which arises from a mismatch between the model for $C_{\ell}^{\dot{\tau} g}$ and the true cross-spectrum (see e.g. Refs.~\cite{Battaglia:2016xbi,Smith:2018bpn,Madhavacheril:2019buy,Giri:2020pkk,Cayuso:2021ljq}). 

Our analysis uses the reconstructions of Ref.~\cite{Bloch:2024} based on the \textit{Planck} PR3~\cite{PlanckCollaboration2020} \texttt{SMICA} and \texttt{Commander} component-separated temperature maps~\cite{Ade2014} and the unWISE~\cite{2010AJ....140.1868W,Mainzer2014,Schlafly_2019} `blue sample'~\cite{Krolewski2020} with a binary foreground mask  that retains $58\%$ of the sky. The plausible range of $b_v$ consistent with modeling choices was found to be $0.5 < b_v < 1.1$~\footnote{This range encapsulates uncertainties in the unWISE blue sample redshift distribution, the degree and scale of small-scale suppression of power in the distribution of baryons due to feedback, and varying assumptions about Helium reionization~\cite{Bloch:2024}.}. To mitigate systematics associated with the choice of component separation technique, we follow Ref.~\cite{Bloch:2024} and use the cross-power of the \texttt{SMICA} and \texttt{Commander}-based reconstructions (\texttt{SxC}) where relevant.  A summary of our analysis choices can be found in the SM. While no statistically significant detection of a signal was made, these data products are still powerful tools to set limits on various cosmological scenarios beyond $\Lambda$CDM.

The angular power spectra of the estimator and galaxy density are given by: 
\begin{equation}\label{eq:spectra}
    C_\ell^{XY}= 4 \pi \int\frac{\dd k}{k} \Delta^{X}_\ell(k)\Delta^{Y}_\ell(k)\mathcal{P}(k) + N^{XY}\,,
\end{equation}
where $XY\in\{\hat{v} \hat{v},\hat{v}g,gg\}$ and  $\mathcal{P}(k)$ is the dimensionless primordial power spectrum. The signal components are determined by the transfer functions $\Delta_\ell^X(k)$. The galaxy transfer function is given by 
\begin{equation}\label{eq:galaxy_transfer}
    \Delta^{g}_\ell(k) \equiv \int\dd\chi\, W_g(k,\chi) \mathcal{S}^m(k,\chi)j_\ell(k\chi)\,,
\end{equation}
where $\mathcal{S}^{m}$ is the source function for matter density computed using CAMB \cite{Howlett:2012mh, Lewis:1999bs}, $W_g(k,\chi)=b(\chi,k) \frac{\dd N}{\dd z}H(\chi)$ is the galaxy window function, $b(\chi,k)$ is the (possibly scale-dependent) galaxy bias, $dN/dz$ is the normalized unWISE blue sample redshift distribution from~Ref.\cite{Krolewski2020}, and $j_\ell$ is the spherical Bessel function. The dipole field transfer function is 
\begin{equation}
    \begin{split}
    \Delta_\ell^{\hat{v}}(k)
    \equiv b_v &\int \dd\chi \   W_v(k,\chi) S^v(k,\chi) \\ 
    &\times [\ell j_{\ell-1}(k \chi) - (\ell+1) j_{\ell+1}(k \chi)]\,, 
    \end{split}
\end{equation}
where $\mathcal{S}^v(k,\chi)$ is the source function for the remote dipole field; further details can be found in the SM. The noise contributions are $N^{\hat{v} \hat{v}} = N$, $N^{gg} = 1/\bar{n}_g$, and $N^{\hat{v} g} = 0$ where $\bar{n}_g = 1.1 \times 10^7$ is the number of galaxies per steradian in the unWISE blue sample.

\noindent \textbf{\textit{Constraints on homogeneity:}}
Most cosmological observables originate on the surface of our PLC, and strictly speaking can only probe the isotropy of the Universe. The remote dipole field is a {\em non-local} observable that depends on the PLC originating from each location, making it a powerful probe of the {\em homogeneity} of the Universe. A number of previous studies have focused on using the kSZ {\em auto-power spectrum} to constrain inhomogeneous but isotropic cosmologies where we inhabit the center of a void (a local under-density)~\cite{Goodman_1995,Caldwell_2008,GBH:2008,Garc_a_Bellido_2008,Biswas_2010,Moss_2011a,Zibin_2011,void:Zhang:2011,Hoscheit:2018,Garfinkle_2010,void:Camarena:2022iae,void:Ding:2019mmw}. An analysis including cross-correlation with LSS that improved on these results was performed in Ref.~\cite{Planck:2013rgv}. Here, we illustrate that the monopole (and low-$\ell$ moments) of the reconstructed remote dipole field can improve on these existing results and provide a powerful probe of inhomogeneous cosmologies.

The reconstruction monopole is sourced by remote dipoles coherently aligned with the line-of-sight. Within $\Lambda$CDM, Eq.~\eqref{eq:spectra} yields a prediction for the signal component of $C_0^{\hat{v} \hat{v}} = 1.73 b_v^2  \times 10^{3} \ {\rm (km/s)^2}$, arising from a chance over- or under-density, comparable in amplitude to the reconstruction noise. We perform a search for a signal monopole $a_{0}^{\hat{v}}$ from the Gaussian likelihood for the \texttt{SxC} reconstruction monopole $\hat{a}_0^{\hat{v}}$ (estimated as the mean of un-masked pixels). The reconstruction monopole is biased by correlated foreground residuals contributing to $C_\ell^{Tg; \rm obs}$~\cite{Bloch:2024}, so we marginalize with uniform priors over a foreground contribution  $a_{0}^{\rm FG}$ in the range $0\leq |a_0^{\rm FG}| \leq 2|\hat{a}_0^{\hat{v}}|$; we additionally marginalize over an optical depth bias in the range $0.5 \leq b_v \leq 1.1$. See the SM for further details. From this posterior, we set the limit $|\langle\hat{v}\rangle| = |a_0^{\hat{v}}|/\sqrt{4\pi} < 38 \ {\rm km/s}$ ($67 \ {\rm km/s}$) at $68\%$ ($95\%$). This limit is roughly a factor of 3 greater than the expected $\Lambda$CDM value, implying stringent constraints on additional sources of large-scale inhomogeneity.

To illustrate the constraints possible for a specific model, we focus on spherically-symmetric void cosmologies described by the commonly-used 
Constrained Garc\'ia-Bellido Haugb\o lle (CGBH) parameterization \cite{GBH:2008}:
\begin{equation}\label{GBH:2008}
    \delta(\chi,z=0)=\delta_V \left(\frac{1-\mathrm{tanh}[(\chi-r_V)/2\Delta r]}{1+\mathrm{tanh}(r_V/2\Delta r)}\right)\,,
\end{equation}
where $\delta$ is the synchronous-gauge total matter perturbation at $z=0$, $\delta_V$ is the depth of the void, $r_V$ is the radius of the void, and $\Delta r$ is the steepness of the void edge. Given the small observed reconstruction monopole, we work within linear theory (see e.g. Ref.~\cite{Van_Acoleyen_2008} for the mapping between the non-linear Lema\^{i}tre-Tolman-Bondi (LTB) solutions and linear theory). We compute the expected monopole Eq.~\eqref{eq:estim_mean} using {\sc CAMB} to find the Newtonian-gauge peculiar velocity along the PLC of an observer at the center of the void from Eq.~\eqref{GBH:2008}; see the SM for the full computation.

\begin{figure}
\includegraphics[width=.8\linewidth]{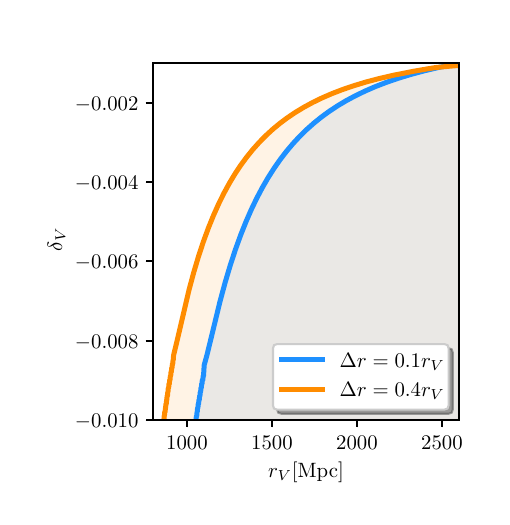}
         \label{fig:commvoid}
         \vspace*{-1cm}
    \caption{The $68\%$ constraints from the \texttt{SMICA}-\texttt{Commander}-reconstruction monopole for an observer at the center of a CGBH void Eq.~\eqref{GBH:2008}. The shaded region shows the ruled out values of depth ($\delta_V$) and width ($r_V$), for a given steepness profile ($\Delta r$).}
    \label{fig:void}\vspace*{-0.5cm}
\end{figure}

Modeling the observed \texttt{SxC} reconstruction monopole as reconstruction noise, $\Lambda$CDM inhomogeneities, foregrounds and the contribution from the void, we compute the posterior over the void depth $\delta_V$ and size $r_V$ at two widths $\Delta r$. The $68\%$ upper-limits on the void parameters obtained after marginalizing over optical depth bias and foreground amplitude are shown in Fig.~\ref{fig:void}. 
If we are not at the center of the void, there will additionally be a reconstruction dipole, which yields tighter constraints; see the SM for further details. It is straightforward to extend these results to other void models, or other sources of large-scale inhomogeneity such as cosmic bubble collisions~\cite{Aguirre:2007an,Zhang:2015uta}. We expect any such model to be constrained near the level of $\Lambda$CDM fluctuations.

\noindent \textbf{\textit{Probing the rest-frame of LSS:}} Within $\Lambda$CDM, adiabatic initial conditions guarantee that the CMB and LSS share a common rest frame. This agreement has been tested by comparing our locally observed CMB dipole $v=369.82 \pm 0.11 \ {\rm km/s}$ in the direction $\{\ell,b\}  = \{264.021\pm0.011^\circ , 48.253\pm0.005^\circ \}$~\cite{PlanckCollaboration2020} to the dipole in the observed number counts of various tracers of LSS~\cite{EllisBaldwin1984,Secrest:2022uvx,Wagenveld:2023kvi,Darling:2022jxt,Murray:2021frz,Rubart:2013tx,Sorrenti:2022zat,Horstmann:2021jjg,Secrest:2022uvx,2021ApJ...908L..51S,Dam:2022wwh,Peebles:2022akh,cheng2024radiosourcedipolenvss,Abghari:2024eja,Gibelyou:2012ri}, the aberration of the small-scale CMB anisotropies~\cite{Challinor_2002,Saha:2021bay,Planck:2013kqc}, and the modulation of SZ effects~\cite{Planck:2013rgv,Planck:2020qil}. The quoted precision of each of these methods is roughly $\mathcal{O}(100) \ {\rm km/s}$. For galaxy number count dipoles, some studies find a result in tension with the CMB dipole (up to 5.1$\sigma$ \cite{Secrest:2022uvx}); see e.g. Refs.~\cite{Peebles:2022akh,cheng2024radiosourcedipolenvss,Abghari:2024eja} for an assessment. A confirmed discrepancy between the CMB dipole and these other probes could be interpreted as evidence for an `intrinsic' dipole, a fundamental difference in the rest-frame of the CMB and LSS. Here, we demonstrate that the intrinsic dipole can be constrained using the $\ell = 1$ components of the remote dipole field~\cite{Deutsch2018a,Cayuso2018} at a sensitivity exceeding that of other probes.

We model the observed reconstruction dipole as the sum of reconstruction noise, the expected $\Lambda$CDM signal $C_1^{\hat{v}\hat{v}} = 4.76 b_v^2 \times 10^{2} \ {\rm (km/s)^2}$, and an unknown intrinsic dipole $v^{\rm int} (\nhat) = D^{\rm int} \ Y_{10}(\nhat-\nhat^{\rm int})$ with amplitude $D^{\rm int}$ oriented at an angle $\nhat^{\rm int}$. We perform a search for an intrinsic dipole by estimating the full-sky reconstruction multipoles using a quadratic maximum likelihood (QML) estimator~\cite{ Tegmark_1997} as implemented in Ref.~\cite{Vanneste:2018azc} on the \texttt{SxC} reconstructions and constructing a posterior for $D^{\rm int}$ from the Gaussian likelihood for the $\ell=1$ estimated reconstruction multipoles on the full sky~\footnote{We validate this approach by comparing against a pixel-space likelihood in the SM, finding reasonable agreement.}. Marginalizing over the orientation of the intrinsic dipole and the optical depth bias, we bound $D^{\rm int} < 14 \ {\rm km/s}$ ($26 \ {\rm km/s}$) at $68\%$ confidence ($95\%$). This greatly exceeds the sensitivity of existing techniques to the magnitude of an intrinsic dipole. 

This generic constraint can be extended to particular models. For example, a large intrinsic dipole can be sourced by a superhorizon matter-radiation isocurvature mode~\cite{Turner1991,Langlois:1995ca}~\footnote{Diffeomorphism invariance leads to a suppression of the intrinsic dipole from superhorizon adiabatic modes~\cite{Turner1991,Erickcek:2008jp,Zibin:2008fe,Mirbabayi:2014hda}, a statement that extends to the remote dipole field~\cite{Zhang:2015uta,Terrana2017}.} while remaining consistent with the observed CMB quadrupole~\cite{Turner1991,Erickcek:2008jp,Erickcek2009}. With sufficient fine-tuning, such modes could address discrepancies between the CMB and LSS dipole~\cite{Domenech:2022mvt}. We parameterize a single superhorizon matter-radiation isocurvature mode by an amplitude $A_{m\gamma}$, wavenumber $\tilde{k}$, and phase $\alpha$ (where $\alpha=0$ corresponds to positioning us at a node). Orienting the mode along the $z$-axis, we have $a_{10}^{\hat{v}} \propto  A_{m\gamma} \tilde{k} \chi_{\rm dec} \cos \alpha$. The same mode also contributes to the CMB quadrupole as $a_{20}^\Theta \propto A_{m\gamma} (\tilde{k} \chi_{\rm dec})^2 \sin \alpha$. The CMB quadrupole is suppressed by an additional power of  $(\tilde{k} \chi_{\rm dec}) \ll 1$, implying that unless $\alpha \simeq \pi/2$, the dipole field is a more sensitive probe of superhorizon isocurvature. 

We construct the posterior over the parameter combination $(A_{m\gamma} \tilde{k} \chi_{\rm dec})$ from the Gaussian likelihood for the \texttt{SxC} reconstruction. Marginalizing over $\alpha$, orientation of the isocurvature mode, and the optical depth bias we bound $(A_{m\gamma} \tilde{k} \chi_{\rm dec}) < 6.90 \times 10^{-4}$ ($2.04 \times 10^{-3}$) at $68\%$ ($95\%$) confidence. We can interpret this bound as ruling out non-linear modes ($A_{m\gamma} \sim 1$) with wavelength less than $\sim 10^3 \chi_{\rm dec}$ or ruling out modes with $\tilde{k}\chi \sim 1$ that have an amplitude greater than $\sim 10^{-3}$. This bound strongly rules out matter-radiation isocurvature as the source of the discrepancy between the CMB and LSS dipoles~\cite{Domenech:2022mvt}, and could have implications for models of the observed hemispherical power asymmetry in the primary CMB~\cite{Erickcek:2008jp,Erickcek2009,Liddle:2013czu,Dai:2013kfa}.

\noindent \textbf{\textit{Constraining primordial non-Gaussianity and compensated isocurvature:}} An important goal of on-going CMB and LSS experiments is to distinguish between classes of inflationary models with a single or multiple degrees of freedom by searching for primordial non-Gaussianity (PNG) and isocurvature. Local-type PNG, which couples short- and long-wavelength modes, and dark matter-baryon isocurvature (compensated isocurvature perturbations~\cite{Gordon_2009,Holder_2010} or `CIP's), induce a scale-dependent galaxy bias~\citep{Dalal:2007cu,Barreira:2019qdl}. A powerful method to measure this scale-dependent bias is through the cross-correlation of the kSZ velocity reconstruction with a galaxy survey~\cite{Munchmeyer2018,Contreras:2019bxy,Hotinli:2019wdp}, which in the high signal-to-noise regime can take advantage of sample variance cancellation~\cite{Seljak_2009}. Here, we set the stage for these future results by computing the constraints on PNG and CIPs from the \textit{Planck}-unWISE reconstruction correlated with unWISE.
 
The scale-dependent bias for both PNG and CIPs (we focus on the case where CIPs trace the primordial curvature perturbation, known as `correlated' CIPs) can be incorporated into the galaxy window function Eq.~\eqref{eq:galaxy_transfer} as
\begin{eqnarray}\label{eq:nongawinf}
{W}_g(\chi(z))&=&\left[b_{\rm G}(z)+b_{\rm NG}(k,z)\,f_{\rm NL}\right]\frac{\dd N}{\dd z}H(z), \\
{W}_g(\chi(z))&=&\left[b_{\rm G}(z)+b_{\rm CIP}(k,z)A_{\rm CIP}\right]\frac{\dd N}{\dd z}H(z)
\end{eqnarray}
where $b_{\rm G}(z)=0.8+1.2 z$ is the estimated unWISE blue galaxy bias from Ref.~\cite{Krolewski2020}, $f_{\rm NL}$ is the amplitude of PNG, $A_{\rm CIP}$ is the amplitude of CIPs, and the functions $b_{\rm NG}(k,z)$ and $b_{\rm CIP}(k,z)$ encode the redshift and scale-dependence ($\propto k^{-2}$ in both cases). We fix these functions in our analysis as computed in Refs.~\cite{Contreras:2019bxy,Hotinli:2019wdp}; see the SM for a summary. The current sensitivity to $f_{\rm NL}$ from scale-dependent bias is  $\sigma_{f_{\rm NL}} \sim 30$ from eBOSS data~\cite{eBOSS:2021jbt,eBOSS:2019sma,Cabass:2022ymb,DAmico:2022gki}, some distance from the critical threshold for probing multi-field inflation $\sigma_{f_{\rm NL}} \lesssim 1$. For correlated CIPs the sensitivity from scale-dependent bias is $\sigma_{A_{\rm CIP}} \sim 60$ (assuming fixed $b_{\rm CIP}(k,z)$ as we do here) using BOSS DR12~\cite{Barreira:2023uvp}. Interesting thresholds for CIPs are $A_{\rm CIP}=16$ if baryon (CDM) number is produced by (before) curvaton decay and $A_{\rm CIP}=-3$ if CDM (baryon) number is produced by (before) curvaton decay.

\begin{figure}[t!]
    \includegraphics[width=0.49\textwidth]{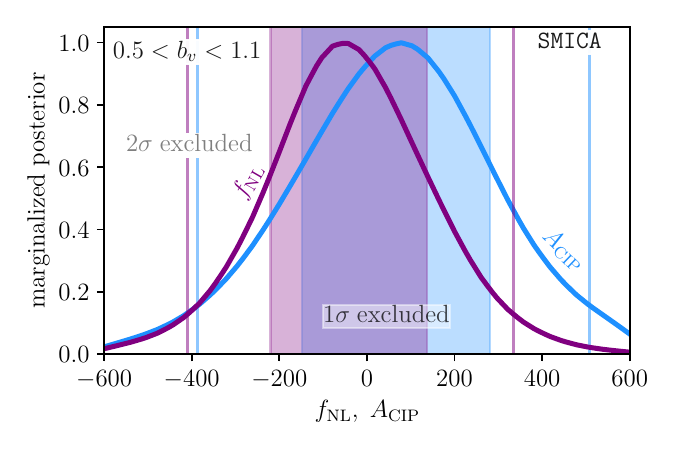} 
    \vspace*{-1cm} 
    \caption{1-dimensional marginalized posteriors on $f_{\rm NL}$ and $A_{\rm CIP}$ from \texttt{SMICA}. We assume a uniform prior on the velocity bias within the range $b_v\in[0.5,1.1]$. 
    } \label{fig:fNL1}
    \vspace*{-0.3cm}
    
\end{figure}

We perform a search for PNG and CIPs using the reconstruction-galaxy cross-power $\hat{C}_{\ell}^{\hat{v}g}$ alone since the reconstruction is noise-dominated, and the unWISE blue sample is heavily contaminated by systematics on large-angular scales~\footnote{Note that this means we do not benefit from the boost in sensitivity from sample variance cancellation that will be possible with future datasets.}. To contend with the mode-coupling induced by the reconstruction mask, we calculate spectra by again using the QML estimator; see the SM for further details.

We construct the posterior over the parameter sets $\{b_{v},f_{\rm NL}\}$ and $\{b_{v},A_{\rm CIP}\}$ at fixed cosmological parameters and galaxy bias; marginalizing over these parameters would weaken constraints. The likelihood function is evaluated by brute force on a 40-by-30 grid of $f_{\rm NL}$ (or $A_{\rm CIP}$) and $b_v$ values by creating $10^4$ masked Gaussian simulations produced from the theory spectra $\{C_\ell^{\hat{v}\hat{v}},C_\ell^{\hat{v}g},C_\ell^{gg}\}$ and using the QML to measure $\hat{C}_{\ell}^{\hat{v}g}$ for each realization. The sampled parameter range is $\in[-600,600]$ for both $A_{\rm CIP}$ and $f_{\rm NL}$; and $0<b_v<6$; we assume uniform priors over all parameters to obtain the normalized posterior including multipoles $\ell\in[2,20]$.

Our marginalized posteriors on $f_{\rm NL}$ and $A_{\rm CIP}$ are shown in Fig.~\ref{fig:fNL1}. Here we have taken a uniform prior in $0.5\leq b_v \leq 1.1$ and used the \texttt{SMICA}-based reconstruction. The $68$ ($95$) percentile excluded region correspond to $-220\lesssim f_{\rm NL}\lesssim 136$ and shown by the shaded purple region ($-409\lesssim f_{\rm NL}\lesssim 335$ and shown by purple vertical lines) and $-147\lesssim A_{\rm CIP} \lesssim 281$ ($-384 \lesssim A_{\rm CIP} \lesssim 509$) shown in blue. The full results for both \texttt{SMICA} and \texttt{Commander}-based reconstructions are recorded in Tables~\ref{tab:fnl_acip_b1}~and~\ref{tab:fnl_acip_b2} for uniform prior assumptions on $b_v$ within the ranges $[0.5,1.1]$ and $b_v<4$, respectively. Assuming the wider range of $b_v$ weakens the constraints by $\mathcal{O}(10\%)$. With the tighter $b_v$ prior, our constraints are roughly a factor of 5 weaker than existing constraints on both scenarios from scale-dependent bias. However, these bounds from kSZ velocity reconstruction will improve rapidly with future datasets.

\noindent \textbf{\textit{Conclusions:}}  We have demonstrated that it is possible to constrain a wide variety of beyond-$\Lambda$CDM cosmologies even in the current noise-dominated regime of kSZ velocity reconstruction. This work can be used as a template for the analysis of data from CMB (e.g. SO) and galaxy surveys (e.g. Dark Energy Spectroscopic Instrument (DESI)\footnote{\url{https://www.desi.lbl.gov/}}) currently in progress, which will rapidly catapult kSZ velocity reconstruction into the signal-dominated regime. The scenarios presented here are only a representative sample of the full-spectrum of possible cosmological constraints we can expect from kSZ velocity reconstruction using these datasets, and give a first-glimpse at the broad program for probing fundamental physics using the secondary CMB.

\section{Acknowledgements}

We thank Richard Bloch, Gil Holder, and Kendrick Smith for useful discussions. We thank Alex Krolewski for sharing expertise and data from the unWISE catalogue. SCH is supported by P.~J.~E.~Peebles Fellowship at the Perimeter Institute for Theoretical Physics. MCJ is supported by the National Science and Engineering Research Council through a Discovery grant. JK acknowledges support from the Natural Sciences and Engineering Research Council of Canada (NSERC) through the Vanier Canada Graduate Scholarship. This research was supported in part by Perimeter Institute for Theoretical Physics. Research at Perimeter Institute is supported by the Government of Canada through the Department of Innovation, Science and Economic Development Canada and by the Province of Ontario through the Ministry of Research, Innovation and Science. 
This project made use of the software tools ReCCO code~\cite{Cayuso:2021ljq}, {\sc CAMB}~\cite{Lewis:1999bs,Howlett:2012mh}, {\sc Numpy}~\cite{numpy}, {\sc healpy}~\cite{healpy}, {\sc Matplotlib}~\cite{matplotlib}, and {\sc SciPy}~\cite{Scipy}. Some of the calculations were also done using the Symmetry computing cluster at Perimeter Institute. 

\bibliography{planck_unwise_cos_refs}


\appendix

\section{kSZ Velocity Reconstruction with \textit{Planck} and unWISE}

In this section we provide a summary of the main results of Ref.~\cite{Bloch:2024}. We utilize the reconstructed remote dipole field produced from the quadratic estimator Eq.~\eqref{eq:quadestim} using the PR3 \texttt{SMICA} and \texttt{Commander} component-separated CMB maps~\cite{Akrami2020} and the unWISE blue sample number counts map described in Ref.~\cite{Krolewski2020}. Post-reconstruction, we apply a binary mask consisting of the union of the mask used in the unWISE analysis of Ref.~\cite{Krolewski2020} (a galactic cut retaining $70\%$ of the sky and a mask removing stars, planetary nebulae, and bright sources) and the \texttt{SMICA}-based confidence mask of Ref.~\cite{Akrami2020}; the full mask retains $f_{\rm sky} = 0.58$ of the sky. The quadratic estimator requires a model for the optical depth-galaxy cross power spectrum. We use the fiducial model of Ref.~\cite{Bloch:2024}, which assumes the unWISE galaxy power spectrum is well-described by a linear bias $b_g = 0.8 + 1.2 z$ (a fit from Ref.~\cite{Krolewski2020}), and that the distribution of electrons is a scale- and redshift-dependent biased tracer of matter with the fiducual parameters of Ref.~\cite{Takahashi2020}. It was estimated in Ref.~\cite{Bloch:2024} that uncertainties in the modeling of the optical depth-galaxy cross power spectrum can plausibly lead to an optical depth bias in the range $0.5 \leq b_v \leq 1.1$. This range is used as a prior in our analysis. We summarize various properties of the reconstructions in Table~\ref{tab:reconstruction_properties}, as described in more detail below.

\begin{table}
    \centering
    \begin{tabular}{|c|c|c|c| }
    \hline
         Map & $N \ (10^{-9})$ & $\hat{a}_0^{\hat{v}} \ (10^{-4})$ & $\hat{C}_1^{\hat{v}\hat{v}} \ (10^{-9})$ \\
         \hline
         \texttt{SMICA} & $7.39$ & $21.4$ & $3.48$ \\
         \texttt{Commander} & $6.21$ & $-1.13$ & $12.9$ \\
         \texttt{SMICA} x \texttt{Commander} (\texttt{SxC}) & $6.11$ & $4.92$ & $3.48$ \\
         \hline
    \end{tabular}
    \caption{Properties of the reconstruction maps. $N$ is the reconstruction noise, $\hat{a}_0^{\hat{v}}$ is the monopole measured from the average of un-masked pixels, and $\hat{C}_1^{\hat{v}\hat{v}}$ is the dipole found from the QML estimator as described in the text.}
    \label{tab:reconstruction_properties}\vspace*{-0.4cm}
\end{table}

The output reconstructions are scaled by a constant factor $\alpha = \{1.118, 0.934\}$ to correct for (unidentified) systematics in the quadratic estimator so that the estimator pre-factor matches the estimator variance; this yields $N = \{7.39 \times 10^{-9}, 6.21 \times 10^{-9} \}$ for the \texttt{SMICA}- and \texttt{Commander}-based reconstructions respectively. We mitigate potential systematics associated with the component separation technique by using the cross-power spectrum of the \texttt{SMICA}- and \texttt{Commander}-based reconstructions where possible. The reconstruction noise associated with the cross-correlation was found to be $N=6.11 \times 10^{-9}$.

Ref.~\cite{Bloch:2024} found that the reconstruction monopole (measured as the average of un-masked pixels) for individual frequency maps is contaminated by foreground residuals~\footnote{Given a measurement of the temperature-galaxy cross-power spectrum $C_{\ell}^{T g, \rm obs}$, the contribution to the estimator variance is
\begin{eqnarray}\label{eq:monpolenoise}
N_{FG} = N^2 \left(\sum\limits_{\ell'}\frac{2\ell'+1}{4\pi}\frac{C_{\ell'}^{T g, \rm obs} C_{\ell'}^{\tau g}}{C_{\ell'}^{g g,\rm obs}C_{\ell'}^{TT,\rm obs}} \right)^2
\end{eqnarray}
}. The \texttt{SMICA} and \texttt{Commander} monopoles are smaller than those for any of the individual frequency maps, consistent with a reduction in foregrounds. However the two monopoles disagree by roughly an order of magnitude.  We would nevertheless like to access the monopole, since as described in more detail below, it contains useful physical information, even within $\Lambda$CDM. To mitigate contaminants, we include an unknown foreground component in our modeling (that in principle could be as large as the observed monopole), and utilize the cross-power of the monopole (e.g. $a_0^{\hat{v}; \rm cross} \equiv |a_0^{\hat{v}; \rm SMICA} a_0^{\hat{v}; \rm Commander}|^{1/2}$) to mitigate the impact of the component separation technique on our results. The monopole for each map is subtracted before subsequent analysis; we record these values in Table~\ref{tab:reconstruction_properties}.

To derive rigorous cosmological constraints from the angular power spectra of the reconstructions and unWISE galaxy density we must contend with the mode-mixing induced by the mask. The reconstructions are expected to be noise dominated, and therefore Ref.~\cite{Bloch:2024} simply re-scaled the cut-sky power spectra $\hat{C}^{\hat{v}\hat{v}; full}_\ell = \hat{C}^{\hat{v}\hat{v}; cut}_\ell/f_{\rm sky}$, as appropriate for a pure white-noise theory power spectrum. However, at the lowest multipoles the reconstruction noise is comparable to the expected signal, and this assumption is insufficient for deriving cosmological constraints. In addition, here we will be interested in the cross-power spectrum between the reconstructions and the unWISE blue galaxy density, where a simple re-scaling of the cut-sky power spectra is not a good estimate of the full-sky cross-power. 

In this work we apply the Quadratic Maximum Likelihood (QML) estimator~\cite{Tegmark_1997} to infer full-sky spectra from the masked reconstructions and galaxy density~\footnote{An alternative approach would be to use a pseudo-$C_\ell$ estimator~\cite{Wandelt:1998qd,Hivon:2001jp,Alonso:2018jzx}, but this is sub-optimal to the QML estimator on large angular scales (see e.g. Ref.~\cite{Efstathiou:2003dj}).}.
This method is based on the construction of a pixel covariance matrix, incorporating both signal and noise components as $\boldsymbol{C} = \boldsymbol{S} + \boldsymbol{N}$ where $\boldsymbol{S} = \sum_i\boldsymbol{P}_\ell C_\ell$ with $\boldsymbol{P}_\ell^{ij} = \partial \boldsymbol{C}^{ij}/\partial C_\ell$ and $\boldsymbol{N}$ is the pixel-noise variance. 

The cross-correlation (xQML) estimator~\citep{Vanneste:2018azc} is designed to optimize the variance reduction in reconstructed power spectra by utilizing the Fisher information matrix, while ensuring the unbiased recovery of the actual power spectra. The power-spectrum estimator satisfies 
\begin{equation}\label{eq:spec_est}
\hat{C}_\ell^{XY}=\sum_\ell [W_{\ell\ell'}]^{-1}\hat{y}_{\ell'}^{XY}\,,\end{equation}
where $X$ and $Y$ superscripts correspond to the velocity reconstruction or galaxy density,
\begin{equation}
    W_{\ell\ell'}=\frac{1}{2}{\rm Tr}[(\boldsymbol{C}^{XX})^{-1}\boldsymbol{P}_\ell(\boldsymbol{C}^{YY})^{-1}\boldsymbol{P}_{\ell'}]\,,
\end{equation}
and $\hat{y}$ is the map-based estimator that satisfies
\begin{equation}
    \hat{y}_\ell^{XY} \equiv (\boldsymbol{d}^X)^T\boldsymbol{E}_\ell\boldsymbol{d}^Y\,,
\end{equation}
with
\begin{equation}\label{eq:spec_est_end}
    \boldsymbol{E}_\ell=\frac{1}{2}(\boldsymbol{C}^{XX})^{-1}\boldsymbol{P}_\ell(\boldsymbol{C}^{YY})^{-1}\,,
\end{equation}
and $\{\boldsymbol{d}^{X}, \boldsymbol{d}^{Y}\}$ are maps of the dipole field reconstruction or unWISE galaxy density. The spectra we input to the xQML for the reconstruction includes the sum of the fiducial $\Lambda$CDM signal described in the next section with $b_v=1$ and the reconstruction noises listed in Table~\ref{tab:reconstruction_properties}. For galaxy density, we use the fiducial theory signal and shot noise for unWISE blue. For the cross-power spectrum, we use the fiducial $\Lambda$CDM signal with $b_v=1$.

\begin{figure*}
    \centering
\includegraphics[width=1.\textwidth]{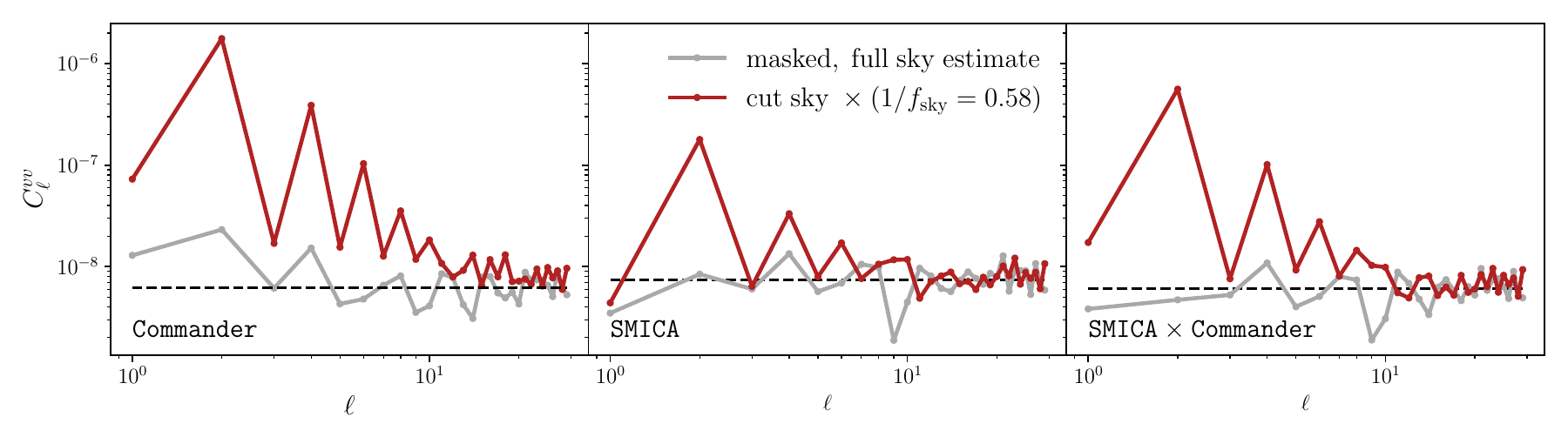} \vspace*{-1cm}
    \caption{The re-scaled cut-sky power spectra (red solid) and Quadratic Maximum Likelihood (gray solid) estimates for the full-sky reconstruction power spectra  compared with the reconstruction noise (black dashed). Values are reported in units of $(v/c)^2$. The re-scaled cut-sky power spectra have clear residuals from the mask on large angular scales, which are reduced by the QML estimate. The two estimates agree with each other and are consistent with the reconstruction noise on small angular scales. Note that the \texttt{Commander}-based reconstruction has significantly more power on large angular scales than the \texttt{SMICA}-based reconstruction.  
    } \label{fig:qmlrecspec}\vspace*{-0.4cm}
\end{figure*}

\begin{figure}[b!]
    \centering
\includegraphics[width=0.475\textwidth]{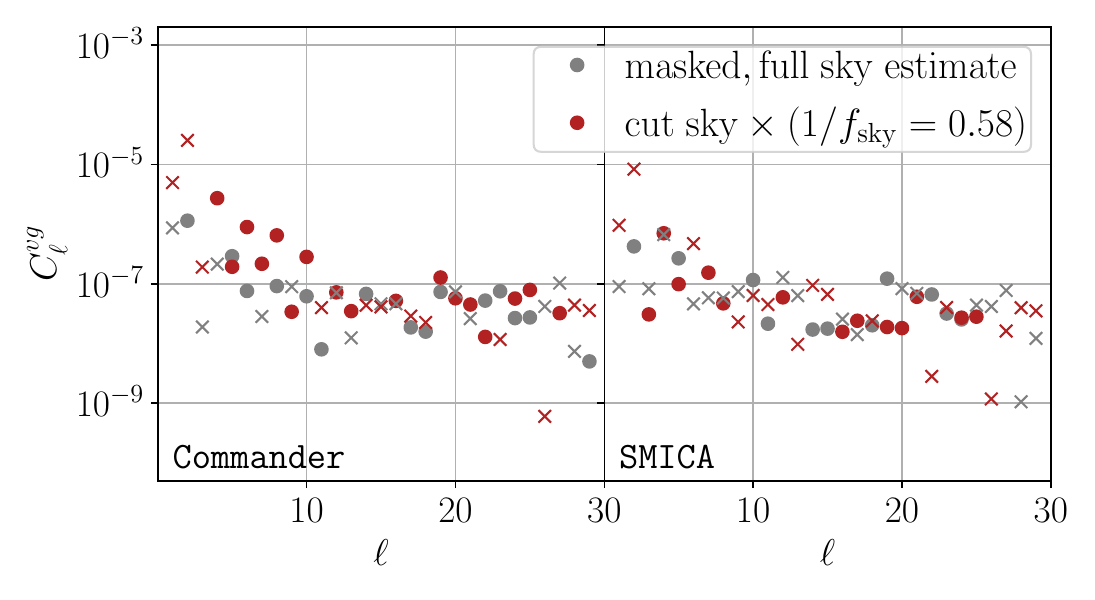} 
    \vspace*{-0.5cm}
    \caption{The re-scaled cut-sky (red) and Quadratic Maximum Likelihood (gray) estimates for the full-sky reconstruction-galaxy density cross-spectra; positive values are indicated by solid circles, negative values are indicated by crosses. Values are reported in units of $(v/c)$. 
    } \label{fig:qmlcrossspec}\vspace*{-0.4cm}
\end{figure}

The xQML estimator can be used to recover the power spectrum at small-$\ell$ from low-resolution maps only. This is not a problem in our analysis, since we are interested in constraining models that yield signatures only at low-$\ell$. Before applying the xQML estimator, we degrade the input maps from $N_{\rm side} = 2048$ to $N_{\rm side} = 16$ (corresponding to pixels with an area of roughly $13 \ {\rm deg}^2$, and a maximum multipole of $\ell_{\rm max} = 48$). We then apply the mask degraded to the same resolution. The resulting spectra for the remote dipole field reconstructions are shown in Fig.~\ref{fig:qmlrecspec} and the dipole field-galaxy cross power is shown in Fig.~\ref{fig:qmlcrossspec}. The reconstruction auto- and cross-power spectra are consistent with the reconstruction noise in Table~\ref{tab:reconstruction_properties}. We also record the estimates for $\hat{C}_1$ in Table~\ref{tab:reconstruction_properties} for reference, since these are used in subsequent analyses.

\section{$\Lambda$CDM Signal}

The remote dipole signal is computed from 
\begin{equation}\label{eq:rdsignal}
    v(\nhat) = \frac{3}{4 \pi} \int d\chi\, W_v(\chi) \int d^2\nhat_{\rm dec} \, \Theta (\nhat, \chi, \nhat_{\rm dec}) (\nhat \cdot \nhat_{\rm dec})
\end{equation}
where $W_v(\chi)$ is the galaxy window function, $\nhat$ is our line of sight, $\nhat_{\rm dec}$ is a line of sight from the point $\chi \nhat$, and $\Theta (\nhat, \chi, \nhat_{\rm dec})$ is the temperature field. The temperature field receives contributions from the Sachs-Wolfe, Doppler, and integrated Sachs-Wolfe effects
\begin{eqnarray}
    \Theta(\nhat, \chi, \nhat_{\rm dec}) &=& \Theta_{\rm SW} (\nhat, \chi, \nhat_{\rm dec})  \\
    &+& \Theta_{\rm ISW} (\nhat, \chi, \nhat_{\rm dec})  + \Theta_{\rm D} (\nhat, \chi, \nhat_{\rm dec}) \nonumber
\end{eqnarray}
We work in Newtonian gauge for adiabatic initial conditions with metric
\begin{equation}
ds^2 = -(1+2 \Psi) dt^2 + a^2(t) (1-2\Psi) dx^2     
\end{equation}
In the limit of instantaneous recombination and matter domination, the contributions to the temperature field are 
\begin{eqnarray}\label{eq:Thetas}
    \Theta_{\rm SW} &=& \frac{1}{3} \Psi[(\chi_{\rm dec} - \chi)\nhat_{\rm dec}] \nonumber \\
    \Theta_{\rm ISW} &=& 2 \int_{\chi_{\rm dec}}^{\chi} d\chi \ \frac{d}{d\chi} \Psi[\chi \nhat+(\chi_{\rm dec} - \chi)\nhat_{\rm dec}] \\
    \Theta_{\rm D} &=& \nhat_{\rm dec} \cdot \left[ {\bf v}(\chi,\nhat) - {\bf v} (\chi_{\rm dec},\nhat_{\rm dec}) \right] \nonumber
\end{eqnarray}
where $\chi_{\rm dec}$ is the radial comoving distance to decoupling from our position. 

The multipole moments of the remote dipole field are determined by the Fourier components of the primordial Newtonian potential $\Psi_0 ({\bf k})$ by
\begin{eqnarray}\label{eq:almv}
    a_{\ell m}^v = 4 \pi (-i)^\ell \int \frac{d^3k}{(2\pi)^3} \ \Delta^{v}_\ell(k) \Psi_0 ({\bf k}) Y_{\ell m}^* (\khat) 
\end{eqnarray}
where
\begin{eqnarray}\label{eq:lcdm_transfer}
    \Delta_\ell^{v}(k)
    &\equiv& \int \dd\chi \   W_v(\chi) \mathcal{S}^v(k,\chi) \nonumber \\ 
    &\times& [\ell j_{\ell-1}(k \chi) - (\ell+1) j_{\ell+1}(k \chi)]
\end{eqnarray}
The source function $\mathcal{S}^v(k,\chi)$ receives Sachs-Wolfe, Integrated Sachs-Wolfe, and Doppler contributions:
\begin{equation}
    \mathcal{S}^v(k,\chi) = \mathcal{S}_{\rm SW}^v(k,\chi) + \mathcal{S}_{\rm ISW}^v(k,\chi) + \mathcal{S}_{\rm D}^v(k,\chi)
\end{equation}
From Eq.~\eqref{eq:Thetas} these can be approximated as
\begin{eqnarray}\label{eq:adiabatic_v_sources}
    \mathcal{S}_{\rm SW}^v(k,\chi) &=& D_{\Psi} (k,\chi) j_1 (k [\chi_{\rm dec} - \chi]) \nonumber \\
    \mathcal{S}_{\rm ISW}^v(k,\chi) &=& 6 \int_{\chi_{\rm dec}}^{\chi} d\chi \ \frac{d D_{\Psi} (k,\chi)}{d\chi} j_1 (k [\chi_{\rm dec} - \chi]) \\ 
    \mathcal{S}_{\rm D}^v(k,\chi) &=& k D_v (k,\chi_{\rm dec}) j_0 (k [\chi_{\rm dec} - \chi]) \nonumber \\ 
    &-& 2 k D_v (k,\chi_{\rm dec}) j_2 (k [\chi_{\rm dec} - \chi]) -k D_v (k,\chi) \nonumber
\end{eqnarray}
where the growth functions are defined by
$\Psi ({\bf k},\chi) = D_{\Psi} (k,\chi) \Psi_0 ({\bf k})$ and ${\bf v} ({\bf k},\chi) = D_{v} (k,\chi) {\bf k} \Psi_0 ({\bf k})$ where $\Psi_0 ({\bf k})$ is the initial gravitational potential. The angular power spectrum of the remote dipole signal is related to the primordial power spectrum of the Newtonian potential  $\mathcal{P}_{\Psi}(k)$ through 
\begin{equation}\label{eq:clvv_lcdmpower}
    C_\ell^{vv}= 4 \pi \int\frac{\dd k}{k} \Delta^{v}_\ell(k)^2 \mathcal{P}_\Psi (k)
\end{equation}
We use the ReCCO code~\cite{Cayuso:2021ljq} to compute the transfer function Eq.~\eqref{eq:lcdm_transfer} and signal power spectrum Eq.~\eqref{eq:clvv_lcdmpower} assuming the cosmological parameters \{$H_0=67.5$, $\Omega_b h^2=0.022$, $\Omega_c h^2=0.122$, $\sum m_{\nu}=0.06$, $10^9A_s = 2.0$, $n_s=0.965$\} and the fiducial model for $W_v(\chi)$ described below Eq.~(\ref{eq:estim_mean}). For the signal computation relevant to the unWISE blue sample, the dominant contribution to the remote dipole field comes from the local peculiar velocity field, and one can approximate
\begin{eqnarray}
    \mathcal{S}^v(k,\chi) \simeq -k D_v (k,\chi)
\end{eqnarray}

For adiabatic modes in the long-wavelength limit ($k \chi_{\rm dec} \ll 1$), the remote dipole source function scales like $\mathcal{S}^v(k,\chi) \propto [k (\chi_{\rm dec} - \chi)]^3$. This is in contrast to the naive expectation that the locally observed CMB dipole scale like $j_1 [k (\chi_{\rm dec} - \chi)] \sim k (\chi_{\rm dec} - \chi)$ in this limit, which is the leading order contribution from the Sachs-Wolfe, Integrated Saches-Wolfe, and Doppler contributions defined in Eq.~\eqref{eq:adiabatic_v_sources}. The cancellation of the leading-order terms from each of these sources is a consequence of diffeomorphism invariance, since a purely linear Newtonian potential can be removed by a gauge transformation~\cite{Turner1991,Erickcek:2008jp,Zibin:2008fe,Mirbabayi:2014hda}. As a result, for adiabatic modes, the CMB quadrupole is more sensitive to ultra large-scales than the remote dipole field. 

Note that the monopole ($\ell =0$) is sourced by Eq.~\eqref{eq:rdsignal} with  $\nhat_{\rm dec} = \nhat$ so that $\nhat_{\rm dec} \cdot  \nhat = 1$ and the temperature field is evaluated along the line-of-sight $\Theta (\nhat, \chi, \nhat_{\rm dec} = \nhat)$. The ensemble-average monopole within $\Lambda$CDM is zero; the variance is given by Eq.~\eqref{eq:clvv_lcdmpower} with $\ell = 0$. For our assumed cosmology, we obtain $C_0^{\hat{v} \hat{v}} = 1.93 b_v^2 \times 10^{-8}$ in units of $(v/c)$ or $C_0^{\hat{v} \hat{v}} = 1.73 b_v^2 \times 10^{3} \ {\rm (km/s)^2}$.

\section{General posteriors for the reconstruction monopole and dipole signal}

In this section, we discuss the posterior probability distributions for the reconstruction monopole and dipole used in the main text. On large angular scales we expect that the spherical harmonic coefficients of the signal and noise contributions to the reconstructed remote dipole field are drawn from a Gaussian distribution. The angular power spectrum of the remote dipole field on large angular scales (the average of the squares of spherical harmonic coefficients at fixed multipole $\ell$) is drawn from a Gamma distribution. We can use these simple likelihood functions to build posteriors for a variety of models.

\textbf{\textit{Reconstruction monopole:}} We first find the posterior over the amplitude of signal contributions to the remote dipole field reconstruction monopole. Since we only have access to the un-masked parts of the maps, we estimate the full-sky monopole as the average of the un-masked pixels. Modeling the monopole as the sum of a signal and scale-invariant reconstruction noise, the average of un-masked pixels is an unbiased estimate of the signal contribution. The variance is a factor of $1/f_{\rm sky}$ larger than the full-sky variance set by the reconstruction noise. The likelihood function is therefore
 \begin{eqnarray}\label{eq:monolike}
    P(\hat{a}_0^{\hat{v}} |a_0^{v} ) &\propto& \exp \left[-\frac{\left(\hat{a}_0^{\hat{v}} - b_v a_0^{v} - a_0^{\rm FG} \right)^2}{2 N/f_{\rm sky}} \right]
\end{eqnarray}
In this expression, $a_0^{v}$ is the signal amplitude (including all contributions), $\hat{a}_0^{\hat{v}}$ is the measured monopole, $a_0^{\rm FG}$ is an unknown foreground component, $N$ is the reconstruction noise, and $b_v$ is the optical depth bias. 

To obtain the posterior over a signal $a_0^{v}$, we chose a flat prior $P(b_v)$ for the optical depth bias over the range $0.5 \leq b_v \leq 1.1$ and a flat prior $P(a_0^{\rm FG})$ for the foreground contribution over the range $0 \leq |a_0^{\rm FG}| \leq 2 |\hat{a}_0^{\hat{v}}|$. The prior over foregrounds is chosen to yield a posterior for $a_0^{v}$ with zero mean, with a range large enough to accommodate the hypothesis that the observed monopole is entirely due to an underlying signal or the hypothesis that there are (partially canceling) large foreground and large signal contributions. Note that the observed monopole (see Table~\ref{tab:reconstruction_properties}) for \texttt{Commander} is close to the RMS of the reconstruction noise while \texttt{SMICA} is highly discrepant. We attempt to mitigate these varied results by using the \texttt{SMICA}-\texttt{Commander} (\texttt{SxC}) cross-correlation. Marginalizing over the optical depth and foreground component, we have
 \begin{eqnarray}\label{eq:monoposterior}
    P(a_0^{v} | \hat{a}_0^{\hat{v}} ) \propto \int d b_v \ P(b_v)  \int d a_0^{\rm FG} \ P(a_0^{\rm FG}) P(\hat{a}_0^{\hat{v}} |a_0^{v} )
\end{eqnarray}
In this expression, $a_0^{v}$ is the signal amplitude, $\hat{a}_0^{\hat{v}}$ is the measured \texttt{SxC} monopole from Table~\ref{tab:reconstruction_properties}, and $N$ is the reconstruction noise. Integrating the posterior, we obtain the limits $a_0^{v} < 4.5 \times 10^{-4}$ ($8.0 \times 10^{-4}$) at $68\%$ ($95\%$) confidence in units of $(v/c)$. We relate this to the average velocity over the sky through $|\langle\hat{v}\rangle| = |a_0^{v}|/\sqrt{4\pi}$ which yields $|\langle\hat{v}\rangle| < 38 \ {\rm km/s}$ ($67 \ {\rm km/s}$) at $68\%$ ($95\%$).

\textbf{\textit{Reconstruction dipole:}} Assuming a model for the reconstruction dipole consisting of reconstruction noise and a signal contribution (including the $\Lambda$CDM and any beyond-$\Lambda$CDM contributions), the likelihood function for the reconstruction dipole on the full-sky is
\begin{eqnarray}
P(\hat{a}_{1m}^{\hat{v}} | a_{1m}^{\hat{v}}) \propto \exp \left[ - \frac{\sum_{m=-1}^1 |\hat{a}_{1m}^{\hat{v}} -  a_{1m}^{\hat{v}}|^2}{2N}   \right] 
\end{eqnarray}
where $a_{1m}^{\hat{v}}$ are the signal multipoles, $\hat{a}_{1m}^{\hat{v}}$ are the measured multipoles, and $N$ is the reconstruction noise. Considering the three $\ell=1$ multipoles as the elements of a three-vector, we define the magnitudes
\begin{equation}
   \sum_{m=-1}^1  |\hat{a}_{1m}^{\hat{v}}|^2 = 3 \hat{C}_1^{\hat{v} \hat{v}}, \ \ \sum_{m=-1}^1  |a_{1m}^{\hat{v}}|^2 = 3 C_1^{\hat{v} \hat{v}} \ ,
\end{equation}
and the angle $\theta$ between the measured and signal dipoles
\begin{eqnarray}
    \sum_{m=-1}^1  |\hat{a}_{1m}^{\hat{v}} a_{1m}^{\hat{v}}| = 3 \sqrt{\hat{C}_1^{\hat{v} \hat{v}} C_1^{\hat{v} \hat{v}}} \cos \theta \ .
\end{eqnarray}
We then change variables to $C_1^{\hat{v} \hat{v}}$, the angle $\theta$, and an azimuthal angle $\phi$ in the likelihood. Since none of the physical models we consider have a preferred orientation for the signal dipole, we use a prior that weights the dipole direction uniformly in solid angle. We define the posterior by marginalizing over $\theta$ and $\phi$, yielding:
\begin{equation}\label{eq:dipoleposetriorsig}
\begin{split}
 P&(C_1^{\hat{v} \hat{v}}|\hat{C}_1^{\hat{v} \hat{v}})
 \propto  \int d\theta d\phi \sin \theta \\
 &\times\exp \left[ - \frac{3}{2N} \left( \hat{C}_1^{\hat{v} \hat{v}} + C_1^{\hat{v} \hat{v}}\!-\!2 \sqrt{C_1^{\hat{v} \hat{v}} \hat{C}_1^{\hat{v} \hat{v}}} \cos \theta \right) \right]\,. 
\end{split}
\end{equation}
Unlike for the reconstruction monopole, there is no strong evidence for foreground contamination in the dipole~\cite{Bloch:2024}, and therefore we do not include this component in our model. 

The posterior Eq.~\eqref{eq:dipoleposetriorsig} is formulated on the full-sky, while the reconstructions must be masked. Our primary method to handle this complication is to use the QML estimate for the full-sky dipole $\hat{C}_1^{\hat{v} \hat{v}}$ in the likelihood, based on the \texttt{SxC} values in Table~\ref{tab:reconstruction_properties}. To provide an independent check for potential inaccuracies in the QML estimate on the harmonic-space posterior, we formulate an alternative pixel-based likelihood:
\begin{eqnarray}\label{eq:pixelvar}
\begin{split}
P(\hat{a}_{1m}^{\hat{v}} &| \mathbf{d}^X, \mathbf{d}^Y ) \\  & \propto \exp \left[ - \frac{1}{2}\sum_{i,j} (d^X_i -s_i) [\tilde{N}^{XY}_{ij} ]^{-1}(d^Y_j -s_j)  \right]
\end{split}
\end{eqnarray}
where $\mathbf{d}^X$ are the map values at un-masked pixels labeled by $i$, the dipole signal is modeled as
\begin{eqnarray}
    s_i = b_v \sum D_m Y_{1m}(\nhat_i)
\end{eqnarray}
where the coefficients $D_m$ are chosen to have fixed amplitude $D$ (defined by $D^2 = \sum_m |D_m|^2/3$) but random phase (corresponding to a random orientation of the signal dipole). We have also explicitly included the optical depth bias $b_v$. The covariance is
\begin{equation}
  \tilde{N}^{XY}_{ij} = \tilde{N}^{XY} \delta_{ij}   
\end{equation}
where $\tilde{N}^{XY}$ is the pixel variance. Importantly, note that our model for the likelihood function does not account for any signal component besides a dipole. Within $\Lambda$CDM this is the largest amplitude contribution to the remote dipole field (see preceding section), however higher multipoles would also contribute at a sub-leading level to the remote dipole field measured at each un-masked pixel. We explicitly account for the full $\Lambda$CDM signal in the harmonic-space approach. 

We evaluate Eq.~\eqref{eq:pixelvar} by first degrading the input \texttt{SMICA}-based and \texttt{Commander}-based reconstruction maps and the reconstruction mask to $N_{\rm side} = 128$ using the \texttt{healpy} \texttt{ud\_grade} function. We compute the pixel variance from the harmonic-space reconstruction noise $N$ through
\begin{eqnarray}
    \tilde{N}_{ij} &=& N \sum_{\ell}^{\ell_{\rm max}} \frac{2 \ell +1 }{4\pi} P_{\ell} (\cos \theta_{ij}) \nonumber \\
    &\simeq& N \frac{9 N_{\rm side}}{4\pi} \delta_{ij}
\end{eqnarray}
where we've used the band-limited value $\ell_{\rm max} = 3N_{\rm side}-1$. We then evaluate Eq.~\eqref{eq:pixelvar} over a grid of $50$ values of D between $0 < D < 4 \times 10^{-4}$ and 500 random orientations of the dipole for all permutations of \texttt{SMICA} and \texttt{Commander} data vectors. We average the result at each $D$ over the 500 orientations and normalize the resulting distribution to obtain a posterior distribution over $b_v D$. 

\begin{figure}
    \centering
\includegraphics[width=.4\textwidth]{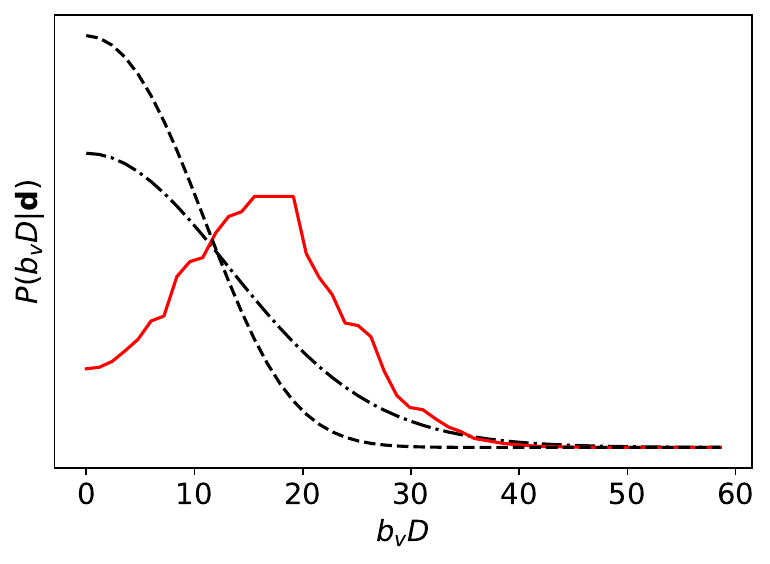} 
    \caption{The pixel-based (red solid) and harmonic-space (black dashed) posterior distributions for a reconstruction dipole signal characterized by an amplitude $D$ marginalized over random orientations. We also plot (black dot-dashed) the constraint on the fundamental dipole that results from marginalizing over the $\Lambda$CDM signal. The three distributions are in broad agreement.} 
     \label{fig:cross_dip_posts}\vspace*{-0.4cm}
\end{figure}

We directly compare the pixel-based posterior with the harmonic-space posterior Eq.~\ref{eq:dipoleposetriorsig} (setting $ C_1^{\hat{v} \hat{v}} = b_V^2 D^2$). The results for posteriors based on the \texttt{SxC} cross-power are shown in Fig.~\ref{fig:cross_dip_posts}. The pixel based posterior using only \texttt{Commander} data prefers a non-zero value of $b_vD = 34.6^{+9.5}_{-7.0} \ {\rm km/s}$, while the posterior using only \texttt{SMICA} prefers $b_v D = 0$~\footnote{In general, \texttt{Commander} appears to have excess power on large angular scales when compared with \texttt{SMICA}. This can be seen in Fig.~\ref{fig:qmlrecspec} and was noted in Ref.~\cite{Bloch:2024}}. This persists in the cross-power based posterior in Fig.~\ref{fig:cross_dip_posts} (red solid curve) as a slight preference for non-zero $b_v D$. The harmonic-space posterior based on the cross-power (black dashed curve) prefers $b_v D = 0$. However, we can see that the two distributions are broadly in agreement. The $68\%$ and $95\%$ exclusion limits for $b_v D$ for all cases are shown in Table~\ref{tab:dipole_limits}. These results produced with a variety of data products and posteriors, make it difficult to accommodate a signal dipole with an amplitude $\agt 30-40 \ {\rm km/s}$.

Given the reasonable agreement between the pixel-based and harmonic-space posteriors found above, we utilize the simpler harmonic-space posteriors in the analysis of the models presented below. 

\begin{table}
    \centering
    \begin{tabular}{|c|c|c|c| }
    \hline
         Data Combination & Posterior & $68\%$ limit (km/s)& $95\%$ \\
         \hline
         \texttt{SMICA} & Pixel & $11$ & $20$ \\
         \texttt{SMICA} & Harmonic & $11$ & $18$ \\
         \texttt{Commander} & Pixel & N/A & N/A \\
         \texttt{Commander} & Harmonic & $18$ & $25$ \\
         \texttt{SxC} & Pixel & $20$ & $30$ \\
         \texttt{SxC} & Harmonic & $11$ & $18$ \\
         \hline
    \end{tabular}
    \caption{The $68\%$ and $95\%$ confidence upper limits on a signal dipole amplitude $b_v D$ using various data combinations and posteriors. The pixel-based posterior is based on Eq.~\eqref{eq:pixelvar} and the harmonic-space posterior is based on Eq.~\eqref{eq:dipoleposetriorsig}. Since the \texttt{Commander} posterior is centered on a non-zero value of $b_vD = 34.6^{+9.5}_{-7.0} \ {\rm km/s}$, we do not report upper limits here.}
    \label{tab:dipole_limits}
\end{table}

\textbf{\textit{Intrinsic dipole:}} A minor extension of the model for the dipole presented above is to separate the intrinsic dipole from a stochastic contribution due to adiabatic modes within $\Lambda$CDM. In this case, we model the observed dipole of the reconstruction as consisting of reconstruction noise, the $\Lambda$CDM contribution expected from Eq.~\eqref{eq:clvv_lcdmpower}, and an intrinsic dipole signal component with amplitude $D^{\rm int}$ and random orientation. We extend the posterior Eq.~\eqref{eq:dipoleposetriorsig} to this situation to obtain
\begin{widetext}
\begin{eqnarray}\label{eq:dipoleposetrior1}
  P(D^{\rm int}|\hat{C}_1^{\hat{v} \hat{v}})\!\propto\!\int\!\!db_v\, P(b_v) \!\!\!\int\!\!d\theta d\phi \sin \theta  \exp\!\left[ - \frac{3}{2(N+b_v^2 C_1^{v v})} \left( \hat{C}_1^{\hat{v} \hat{v}} + b_v^2 {D^{\rm int}}^2\!-\!2 b_v D^{\rm int} \hat{C}_1^{\hat{v} \hat{v}} \cos \theta \right) \right]\,.
\end{eqnarray}
\end{widetext}
where we have marginalized over the optical depth bias with a uniform prior in the range $0.5 \leq b_v \leq 1.1$ and the orientation of the intrinsic dipole.

The resulting normalized posterior (dot-dashed line) is shown in Fig.~\ref{fig:cross_dip_posts} using the \texttt{SxC} cross-power. Due to the increased cosmic variance from $\Lambda$CDM adiabatic modes and the marginalization over $b_v$ the posterior is broader than those for the total signal. The resulting $68\%$ and $95\%$ limits on the intrinsic dipole using the various data combinations are recorded in Table~\ref{tab:intrinsic_dipole_limits}.

\begin{table}
    \centering
    \begin{tabular}{|c|c|c| }
    \hline
         Data Combination & $68\%$ limit (km/s)& $95\%$ \\
         \hline
         \texttt{SMICA} & $14$ & $26$ \\
         \texttt{Commander} & $22$ & $37$ \\
         \texttt{SMICA} x \texttt{Commander} &  $14$ & $26$ \\
         \hline
    \end{tabular}
    \caption{The $68\%$ and $95\%$ confidence upper limits on an intrinsic dipole amplitude $D^{\rm int}$ using various data combinations derived from the posterior Eq.~\ref{eq:dipoleposetrior1}.}
    \label{tab:intrinsic_dipole_limits}\vspace*{-0.4cm}
\end{table}

\section{Void Signal}

We consider the imprint of a large under-density (e.g. a void) on the remote dipole field. In the absence of a particular theoretical model for the origin of such a void, we adopt a common density profile: the Constrained Garc\'ia-Bellido Haugb\o lle (CGBH) parameterization \cite{GBH:2008}. We assume that the void is spherically symmetric and arises from an adiabatic perturbation with a synchronous gauge total matter density profile today $\delta(\chi) = \rho(\chi)/\bar{\rho} - 1$ given by:
\begin{equation}
\delta(\chi,z=0)=\delta_V\left(\frac{1-\mathrm{tanh}[(\chi-r_V)/2\Delta r]}{1+\mathrm{tanh}(r_V/2\Delta r)}\right)
\end{equation}
where $\delta_V$ is the depth of the void, $r_V$ is the radius of the void, and $\Delta r$ is the steepness of the void edge. We further assume that the void is sourced by only the growing mode, implying that in the past the void arose from a time-independent superhorizon perturbation with a strictly smaller amplitude. We do not speculate on the origin of such a structure. Restricting ourselves to voids with $\delta_V \ll 1$, we apply linear theory to determine the evolution of the void. The general case, including $\delta_V \rightarrow -1$ can be treated using the LTB metric (see e.g. Ref.~\citep{Moffat:2016dbd} for a recent discussion on LTB void solutions). As we demonstrate, the sensitivity of the \textit{Planck}-unWISE reconstruction is within the regime of validity of linear theory for a wide range of parameter space, so we are able to avoid the complications of a fully non-linear general relativistic treatment. 

\begin{figure}
    \centering
\includegraphics[width=0.4\textwidth]{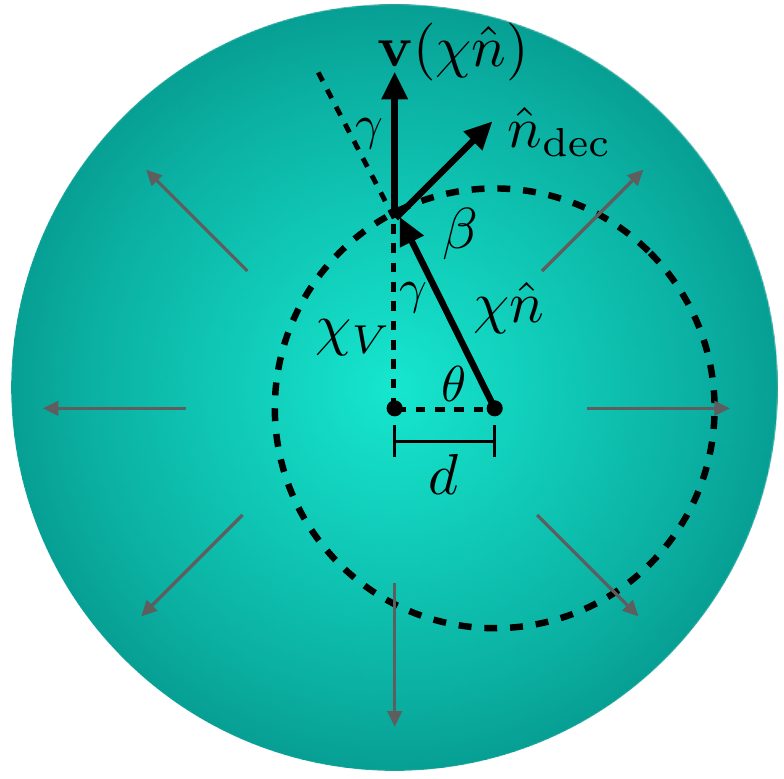} 
    \caption{The void geometry, where the void is shown in green and the observer is located a distance $d$ from the void's center. The void's comoving radial distance is $\chi_V$ and the observer's comoving radial distance is $\chi\nhat$. This means that the velocity induced by the void evaluated at $\chi\nhat$ is $\mathbf{v}(\chi \nhat)$. Then $\nhat_{\rm dec}$ is a line-of-sight from $\chi \nhat$. 
    } \label{fig:void_geometry}\vspace*{-0.4cm}
\end{figure}

Under our assumption that the void is sourced by an adiabatic perturbation, we use the expression Eq.~\eqref{eq:rdsignal} for the remote dipole field with the sources for the temperature field defined in Eq.~\eqref{eq:Thetas}. The local Doppler term is the dominant contribution, and so we approximate 
\begin{eqnarray}\label{eq:void_rd}
        v(\nhat) &\simeq& \frac{3}{4 \pi} \int d\chi\, W_v(\chi) \nonumber \\
        &\times& \int d^2\nhat_{\rm dec} \, (\nhat_{\rm dec} \cdot {\bf v}(\chi,\nhat))  (\nhat \cdot \nhat_{\rm dec})
\end{eqnarray}
Considering an observer located a distance $d$ from the center of the void, the geometry is shown in Fig.~\ref{fig:void_geometry}. First, note the azimuthal symmetry about the line connecting the observer and the center of the void. This allows us to write the dot products in terms of the angles $\gamma$ and $\beta$:
\begin{eqnarray}
    \nhat \cdot \nhat_{\rm dec} &=& \cos \beta \\ 
    \nhat_{\rm dec} \cdot {\bf v}(\chi\nhat) &=& |{\bf v}(\chi\nhat)| (\sin \gamma \sin \beta - \cos \gamma \cos \beta) \\ 
    \cos \gamma &=& \frac{\chi}{\chi_V} \left( 1 - \frac{d}{\chi} \cos \theta \right)
\end{eqnarray}
Integrating over $d^2 \nhat_{\rm dec} = d\phi d\beta \sin \beta$ in Eq.~\eqref{eq:void_rd} we obtain
\begin{eqnarray}\label{eq:void_vphi}
    v(\theta) = - \int d\chi W_v(\chi) \ |{\bf v}(\chi_V)| \frac{\chi}{\chi_V} \left( 1 - \frac{d}{\chi} \cos \theta \right)
\end{eqnarray}
where 
\begin{eqnarray}
    \chi_V^2 = d^2 + \chi^2 -2d\chi \cos\theta 
\end{eqnarray}
and we have explicitly noted the azimuthal symmetry about the axis connecting the observer and the void center by  writing the remote dipole field as a function of $\theta$ only. 

In the limit $d\rightarrow 0$ where the observer is in the center of the void, $\chi_V = \chi$ and the remote dipole field only has a monopole component. The fiducial velocity window function $W_v$ has little support for comoving radial distances $\chi \lesssim 800 \ {\rm Mpc}$ (which encloses only 3\% of $W_v$). One consequence is that if we are at the center of the void, the reconstruction monopole is not sensitive to voids smaller than this. Allowing $d\neq 0$, the remote dipole will vary across the sky. For positions $800 \ {\rm Mpc} \lesssim d \lesssim r_V$, the variation will primarily be in the monopole and dipole moments of $v(\nhat)$. We estimate the monopole and dipole moments by evaluating Eq.~\eqref{eq:void_vphi} along  $\theta = 0, \pi$ and taking:
\begin{eqnarray}\label{eq:mondipvoid}
    a^{v}_{00} \simeq &-\sqrt{4 \pi}& \int d\chi W_v(\chi) \nonumber\\
    &\times&\ \frac{\left(|{\bf v}|(\theta=0) + |{\bf v}|(\theta=\pi) \right)}{2} \\
    a^{v}_{10} \simeq &-\sqrt{\frac{4 \pi}{3}}& \int d\chi W_v(\chi) \nonumber\\
    &\times& \ \frac{\left(|{\bf v}|(\theta=0) - |{\bf v}|(\theta=\pi) \right)}{2} \nonumber
\end{eqnarray}

To evaluate the monopole and dipole moments we find the velocities along the antipodal lines of sight $\theta=0,\pi$ on the past light cone within linear theory. We first describe how this procedure is implemented for a void centered on the origin, and then discuss how it is modified to account for an off-centered observer. We first take the Fourier transform of $\delta(r)|_{z=0}$
\begin{equation}
    \delta(k)|_{z=0} = 4 \pi \int_0^\infty d\chi\, \delta(\chi)|_{z=0} \frac{\mathrm{sin}(k\chi)\chi}{k}
\end{equation}
then use the matter density ($T_\delta (k,z=0)$) and Newtonian-gauge velocity ($T_v (k, z)$) transfer functions from \texttt{CAMB} to evolve the void backwards in time and convert from synchronous-gauge density to Newtonian-gauge velocity:
\begin{equation}
    |{\bf v}|(k,z) = \frac{T_v (k, z)}{T_\delta (k,z=0)} \delta(k)|_{z=0}.
\end{equation}
Next, we inverse Fourier transform the velocity by
\begin{equation}\label{eq:vofchi}
    |{\bf v}|(\chi) = \frac{2}{(2\pi)^2} \int_0^\infty dk\, |{\bf v}| (k,\chi) \frac{\mathrm{sin}(k\chi)k}{\chi}
\end{equation}
and choose the gauge such that far from the void $|{\bf v}|(\chi)$ goes to zero. Then we integrate against the velocity window function $W_v(\chi)$ to obtain the monopole. To compute the velocity along the past light cone of an observer at non-zero $d$ we have to consider separately our two choices of lines of sight, $\theta=0,\pi$. For $\theta=0$, we evaluate Eq.~(\ref{eq:vofchi}) at $d+\chi$, where $\chi$ is the comoving radial distance centered on the void. Then for $\theta=\pi$, we use $\mid d-\chi \mid$ and also need to account for the change in sign of the velocities as the line of sight goes through the middle of the void (ie. $-|{\bf v}|(\mid d-\chi \mid)$ for $\chi\leq d$ and $|{\bf v}|(\mid d-\chi \mid)$ for $\chi > d$). We can then use these to calculate the monopole and dipole using Eq.~(\ref{eq:mondipvoid}).

For the monopole, since the likelihood function for $a_{\ell m}$'s is Gaussian the posterior has the form
 \begin{align}\label{eq:voidmonoposterior}
    P(d_V, r_V | \hat{a}_0^{\hat{v}} )& \propto \int d b_v \ P(b_v)  \int d a_0^{\rm FG} \ P(a_0^{\rm FG}) \nonumber \\ &\times \exp \left[-\frac{\left(\hat{a}_0^{\hat{v}} - b_v d_V a_0^{v}[r_V] - a_0^{\rm FG} \right)^2}{\frac{2}{f_{\rm sky}} (N + b_v^2 C_0^{ v v })}  \right].
\end{align}
Here, our model contains both the signal amplitude $a_0^{v}$ and foreground contributions $a_0^{\rm FG}$, and the data is the measured monopole $\hat{a}_0^{\hat{v}}$ (defined as discussed above by the average of un-masked pixels). The variance includes the reconstruction noise $N$, fraction of the sky $f_{\rm sky}$, and $\Lambda$CDM adiabatic (no void) monopole prediction $C_0^{ v v}$. Our signal $a_0^{v}$ is a function of the void width $r_V$ and since the void depth $d_V$ is just a constant it can be pulled out front. We marginalize over the optical depth bias $b_v$ and the foreground contribution $a_0^{\rm FG}$ with flat priors as described earlier. We use the observed \texttt{SxC} monopole and reconstruction noise given in Table~(\ref{tab:reconstruction_properties}).

The dipole posterior is a specialized version of Eq.~\eqref{eq:dipoleposetrior1}, given by
\begin{widetext}
\begin{align}\label{eq:voiddipoleposetrior1}
 P(d_V, r_V|\hat{C}_1^{\hat{v} \hat{v}})\!&\propto\!\!\int\!\!db_v\, P(b_v) \!\!\!\int\!\!d\theta d\phi \sin \theta  \exp\!\left[ - \frac{3}{2(N+b_v^2 C_1^{v v})} \left( \hat{C}_1^{\hat{v} \hat{v}} + b_v^2 d_V^2 C_1^{v v;V}[r_V]\!-\!2 b_v \sqrt{d_V^2 C_1^{v v;\rm V}[r_V] \hat{C}_1^{\hat{v} \hat{v}}} \cos \theta \right) \right]\,
\end{align}
\end{widetext}
where the void signal is $d_V^2 C_1^{v v;V}[r_V]$, which is a function of $d_V$ and $r_V$, the data is $\hat{C}_1^{\hat{v} \hat{v}}$, and the $\Lambda$CDM adiabatic dipole prediction is $C_1^{v v}$. We again marginalize over the optical depth bias $b_v$ and the angles \{$\theta$, $\phi$\}. For the data, we use the observed \texttt{SxC} dipole and reconstruction noise from Table~(\ref{tab:reconstruction_properties}).

\begin{figure}
\includegraphics[width=.8\linewidth]{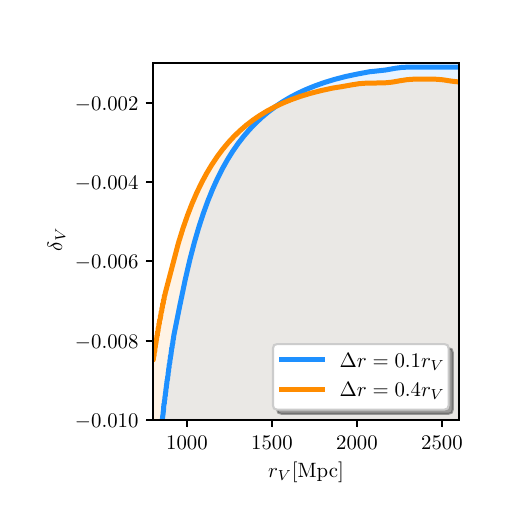}
         \vspace*{-1cm}
    \caption{The \texttt{SxC}-reconstruction monopole and dipole 68\% constraints for an observer off-centered from the CGBH void Eq.~\eqref{GBH:2008} by $d=r_V/2$. The shaded area is the ruled out parameter regions for the void depth ($\delta_V$) and width ($r_V$), for a given steepness profile ($\Delta r$).}
    \label{fig:voidoffcent}\vspace*{-0.4cm}
\end{figure}

In the main text, Fig.~(\ref{fig:void}) shows the parameter regions ruled out by the \texttt{SxC} reconstruction monopole constraints for an observer centered on the void. Here, in Fig.~(\ref{fig:voidoffcent}), are the \texttt{SxC} reconstruction monopole and dipole joint constraints for a void off-centered from the observer by half the distance of the void ($d=r_V/2$). In blue is the steep void profile, with orange as the more gradual profile, and the shaded regions are the discluded void parameters. The off-centered void is more constraining than the centered void because we can use both the reconstruction monopole and dipole. For the monopole alone, the centered and off-centered void constraints are similar -- it is the dipole which performs better than the monopole at constraining the smaller radius, deeper voids, therefore making the joint-constraint tighter. 

\section{Matter-Radiation Isocurvature Signal}

For matter-radiation isocurvature, the remote dipole field signal has the same form as Eq.~\eqref{eq:rdsignal}. However, for isocurvature, the Sachs-Wolfe contribution to the temperature field is
\begin{eqnarray}
    \Theta_{\rm SW}^{\rm iso} = 2  \Psi [(\chi_{\rm dec} - \chi)\nhat_{\rm dec}]
\end{eqnarray}
The primordial matter-radiation isocurvature entropy perturbation $S_{m \gamma} ({\bf k}) \equiv \delta_m - 3\delta_\gamma/4$ sources a potential perturbation on superhorizon scales at early time given by (see e.g.~\cite{Erickcek2009})
\begin{eqnarray}
    \Psi_0 ({\bf k},\chi_{\rm dec}) = - S_{m \gamma} ({\bf k}) / 5
\end{eqnarray}
The multipoles of the remote dipole signal are therefore
\begin{eqnarray}\label{eq:almv_iso}
    a_{\ell m}^v = - 4 \pi (-i)^\ell \int \frac{d^3k}{(2\pi)^3} \ \Delta^{v; {\rm iso}}_\ell(k) S_{m \gamma} ({\bf k}) Y_{\ell m}^* (\khat) 
\end{eqnarray}
with 
\begin{eqnarray}\label{eq:lcdm_transfer_iso}
    \Delta_\ell^{v; {\rm iso}}(k)
    &\equiv& \int \dd\chi \   W_v(\chi) \mathcal{S}^{v; {\rm iso}}(k,\chi) \nonumber \\ 
    &\times& [\ell j_{\ell-1}(k \chi) - (\ell+1) j_{\ell+1}(k \chi)]
\end{eqnarray}
The Integrated Sachs-Wolfe and Dopppler source functions are the same form as the adiabatic modes, however, the Sachs-Wolfe term is relatively larger:
\begin{eqnarray}
    \mathcal{S}_{\rm SW}^{v; {\rm iso}}(k,\chi) &=& \frac{6}{5} \mathcal{S}_{\rm SW}^v(k,\chi), \nonumber \\ \mathcal{S}_{\rm ISW }^{v; {\rm iso}}(k,\chi) &=&\frac{1}{5} \mathcal{S}_{\rm ISW}^v(k,\chi) \\ \mathcal{S}_{\rm D }^{v; {\rm iso}}(k,\chi) &=& \frac{1}{5} \mathcal{S}_{\rm D}^v(k,\chi) \nonumber
\end{eqnarray}
In the long-wavelength limit $k \chi_{\rm dec} \ll 1$, there is a partial cancellation of the components of the source functions (the  same cancellation that occurs for adiabatic modes), and we can make the approximation 
\begin{eqnarray}
    \mathcal{S}^{v; {\rm iso}}(k,\chi) \simeq \mathcal{S}_{\rm SW}^{v}(k,\chi)
\end{eqnarray}
Because of this change in the source-function, the leading-order behavior in the long-wavelength limit is $\mathcal{S}^{v; {\rm iso}}(k,\chi) \sim k[\chi_{\rm dec} - \chi]$. The remote dipole field is therefore far more sensitive to long-wavelength isocurvature than long-wavelength adiabatic modes. 

We consider two scenarios of long-wavelength isocurvature modes. First, we study the case of a single superhorizon isocurvature mode parameterized by
\begin{eqnarray}
    S_{m \gamma} ({\bf x},\chi_{\rm dec}) = A_{m \gamma} \sin \left[\tilde{k} z + \alpha \right]
\end{eqnarray}
where $A_{m \gamma}$ is the amplitude, $\tilde{k}$ is the wavenumber, and the phase $\alpha$ encodes our position (assumed ${\bf x} = 0$) with respect to the node. We have oriented the mode along the z-direction. A short computation yields 
\begin{eqnarray}
    v(\nhat) = - A_{m \gamma} \cos \theta \int &d\chi& \, W_v(\chi) j_1 (k [\chi_{\rm dec} - \chi]) \nonumber \\
    &\times& \cos \left[ \tilde{k} \chi \cos\theta + \alpha \right]
\end{eqnarray}
where $\theta$ is the polar angle, implying that the moments $a^v_{\ell m}$ are non-zero only for $m=0$. Expanding for $\tilde{k} \chi_{\rm dec} \ll 1$, the first moments are
\begin{eqnarray}\label{eq:mrmode_signal}
    a_{10}^v &=& - \sqrt{\frac{4 \pi}{3}} \frac{A_{m \gamma}}{3} \cos \alpha \ \tilde{k} \chi_{\rm dec} \nonumber \\ &\times& \left(1 - \int d\chi\ W_v(\chi) \frac{\chi}{\chi_{\rm dec}} + \mathcal{O}(\tilde{k}^2)   \right) \\
    &\simeq& - 0.57 A_{m \gamma} \tilde{k} \chi_{\rm dec} \cos \alpha \nonumber
\end{eqnarray}
where in the second line we have evaluated the integral using our fiducual velocity window function. The $\ell=0$ and $\ell=2$ moments are both $\mathcal{O}((\tilde{k} \chi_{\rm dec})^2 \sin \alpha)$ and therefore sub-dominant unless $\alpha$ is very close to $\pi/2$. For this reason, we do not perform a joint constraint of the reconstruction dipole with the CMB temperature quadrupole.

The second case we consider is for a scale-invariant long-wavelength stochastic isocurvature background. In analogy with Eq.~\eqref{eq:clvv_lcdmpower}, we have
\begin{eqnarray}
        C_\ell^{vv} = 4 \pi \int\frac{\dd k}{k} \Delta^{v; {\rm iso}}_\ell(k)^2 \mathcal{P}_{\mathcal{S}_{m\gamma}} (k)
\end{eqnarray}
We assume
\begin{eqnarray}
\mathcal{P}_{\mathcal{S}_{m\gamma}} (k) = B_{m \gamma} \Theta(k-k^*)
\end{eqnarray}
where $B_{m \gamma}$ is the constant amplitude of the power spectrum and $\Theta(k-k^*)$ is the Theta function, equal to unity for $k<k^*$. In the long-wavelength limit $k^* \chi_{\rm dec} \ll 1$ the dipole is the dominant multipole
\begin{eqnarray}\label{eq:c1vvisopower}
    C_1^{vv} &\simeq& \frac{2\pi}{9} (k^* \chi_{\rm dec})^2 B_{m \gamma}  \\
    &\times& \left(1 - \int d\chi W_v(\chi) \left[2 \frac{\chi}{\chi_{\rm dec}}-\left(\frac{\chi}{\chi_{\rm dec}}\right)^2 \right]   \right) \nonumber \\
    &\simeq&  0.494  (k^* \chi_{\rm dec})^2 B_{m \gamma}.\nonumber
\end{eqnarray}
In the second line, we have evaluated the integral with our fiducial velocity window function. 

We now describe how we constrain the two scenarios described above. In the first, we wish to use the measured $\ell = 1$ multipole moments to probe new signals {\em at the field level}. That is, we are concerned with the properties of a fixed realization. In the second, we wish to use the measured $\ell = 1$ angular power spectrum to probe the {\em statistical ensemble} from which new signals are drawn. This second scenario includes additional cosmic variance when compared to the first, accounting for the (im)probability of finding a realization of the new signal consistent with observation. 

\textbf{\textit{Single-mode posterior:}} For a single superhorizon matter-radiation isocurvature mode, the posterior from the observed reconstruction dipole is a generalization of Eq.~\ref{eq:dipoleposetrior1} using Eq.~\ref{eq:mrmode_signal} as the dipole amplitude. The free-parameters in the model include the amplitude $A_{m \gamma}$, wave number compared to the distance to decoupling $\tilde{k} \chi_{\rm dec}$, the phase of the mode $\alpha$, and the orientation of the mode. We constrain the parameter combination $(A_{m \gamma} \tilde{k} \chi_{\rm dec})$ by marginalizing over the phase in the range $0\leq\alpha\leq 2 \pi$, the optical depth in the range $0.5 \leq b_v \leq 1.1$, and the orientation of the mode on the sky all with uniform priors. The resulting $68\%$ and $95\%$ upper limits on $(A_{m \gamma} \tilde{k} \chi_{\rm dec})$ using various data combinations are recorded in Table~\ref{tab:mriso_dipole_limits}. 

\begin{table}
    \centering
    \begin{tabular}{|c|c|c| }
    \hline
         Data Combination & $68\%$ limit& $95\%$ \\
         \hline
         \texttt{SMICA} & $7.01 \times 10^{-4}$ & $2.04 \times 10^{-3}$ \\
         \texttt{Commander} & $8.22 \times 10^{-4}$ & $2.09 \times 10^{-3}$ \\
         \texttt{SxC} &  $6.90 \times 10^{-4}$ & $2.04 \times 10^{-3}$ \\
         \hline
    \end{tabular}
    \caption{The $68\%$ and $95\%$ confidence upper limits on the parameter combination $(A_{m \gamma} \tilde{k} \chi_{\rm dec})$ for a super-horizon matter-radiation isocurvature mode using various data combinations. }
    \label{tab:mriso_dipole_limits}\vspace*{-0.4cm}
\end{table}

\textbf{\textit{Stochastic superhorizon posterior:}} In the second scenario, where we wish to constrain the properties of the statistical ensemble, the posterior over the $\ell = 1$ component of the angular power spectrum is 
\begin{equation}
\label{eq:c1vvlikelihood}
\begin{split}
    P(C_1^{vv; {\rm iso}}&|\hat{C}_1^{\hat{v} \hat{v}}) \\
    &\propto\!\int\!db_v\,P(b_v) \left( \frac{ \hat{C}_1^{\hat{v} \hat{v}} }{ b_v^2 (C_1^{vv; {\rm ad}} + C_1^{vv; {\rm iso}})+N }\right)^{3/2} \nonumber \\
    &\times \exp \left[ -\frac{3 \hat{C}_1^{\hat{v} \hat{v}}}{2 (b_v^2 (C_1^{vv; {\rm ad}} + C_1^{vv; {\rm iso}}) +N)}\right] \ .
\end{split}
\end{equation}
Here, $\hat{C}_1^{\hat{v} \hat{v}}$ is the \texttt{SxC} reconstruction dipole from Table~(\ref{tab:reconstruction_properties}). The observed dipole is also modeled as a composition of the reconstruction noise ($N$), the expected adiabatic contribution from Eq.~\eqref{eq:clvv_lcdmpower} ($C_1^{vv; {\rm ad}}$), and an isocurvature signal from Eq.~(\ref{eq:c1vvisopower}) ($C_1^{vv; {\rm iso}}$). We marginalize over $b_v$ as described before, so the posterior is over the parameter combination $(B_{m \gamma} (k^* \chi_{\rm dec})^2)$. The $68\%$ and $95\%$ bounds on this are given in Table~\ref{tab:siso_dip_limits} for the \texttt{SMICA}, \texttt{Commander}, and \texttt{SxC} reconstruction dipoles. 

\begin{table}
    \centering
    \begin{tabular}{|c|c|c| }
    \hline
         Data Combination & $68\%$ limit& $95\%$ \\
         \hline
         \texttt{SMICA} &  $5.30 \times 10^{-7}$ & $2.10 \times 10^{-5}$  \\
         \texttt{Commander} &  $1.00 \times 10^{-6}$ & $3.90 \times 10^{-5}$  \\
         \texttt{SxC} & $5.00 \times 10^{-7}$ & $2.00 \times 10^{-5}$ \\
         \hline
    \end{tabular}
    \caption{The $68\%$ and $95\%$ confidence upper limits on the parameter combination $(B_{m \gamma} (k^* \chi_{\rm dec})^2)$ for a long-wavelength stochastic isocurvature background with a cut-off at mode $k^*$.  }
    \label{tab:siso_dip_limits}\vspace*{-0.4cm}
\end{table}

\section{Correlated Compensated Isocurvature Signal}

Theories that incorporate multiple degrees of freedom can generate isocurvature (entropy) perturbations, where the relative composition of dark matter, baryons, neutrinos, and photons act as independent variables. Although most forms of isocurvature perturbations are stringently limited by current Cosmic Microwave Background (CMB) observations, a significant exception exists: compensated isocurvature perturbations~\cite{Gordon_2009,Holder_2010} (CIPs). CIPs are fluctuations in baryons and cold dark matter that maintain the total matter perturbations as adiabatic and unchanged. These perturbations only affect the CMB through terms that appear at the second order in the matter density contrast~\cite{Grin:2011tf}, making them difficult to constrain. Present measurements from the \textit{Planck} satellite permit an amplitude of CIPs that is approximately 580 times larger than the amplitude of adiabatic modes~\cite{Planck:2018jri}.

Changes in the ratio between baryons and cold dark matter affect the distribution of structure in the Universe, modifying the way galaxies trace the overall matter density. This results in a spatially varying galaxy bias, which connects the observed galaxy over-density to the total matter over-density. Specifically, compensated isocurvature perturbations (CIPs) correlated with primordial curvature perturbations (as seen in scenarios such as the curvaton model) will induce a scale-dependent galaxy bias~\citep{Barreira:2019qdl}. The dominant effect on the scale dependent CIP bias satisfies~\citep{Barreira:2019qdl}
\begin{equation}
b_{\rm CIP}(k,z)=\frac{5 H(z)^2\Omega_m}{2ak^2}f\,b_{\rm CIP}(z)\,,
\end{equation}
where $f=1+{\Omega_b}/{\Omega_c}$. 

We estimate the bias $b_{\rm CIP}(z)$ using the separate Universe approximation by calculating how changes in the baryon-CDM fraction affect the galaxy number density following Ref.~\citep{Hotinli:2019wdp}. For parameters consistent with the unWISE survey blue sample used in this analysis, we find that a quadratic polynomial provides a good fit over the
relevant range of redshifts with 
\be
b_{\rm CIP}(z)\equiv-(0.3 + 0.134\,z + 0.06\,z^2)\,.
\ee
We fix $b_{\rm CIP}(k,z)$ in our analysis, including cosmological parameters and our fit to $b_{\rm CIP}(z)$.

\section{Local-type Primordial Non-Gaussianity Signal}

One of the primary objectives of upcoming large-scale structure surveys and CMB experiments is to detect or tightly constrain primordial non-Gaussianity. Achieving this goal would provide critical insights into the interactions that occurred in the inflationary universe, offering a probe into ultra-high-energy physics that is otherwise inaccessible. Primordial non-Gaussianities have been extensively classified based on their production mechanisms, symmetries, and field content; see e.g. Refs.~\cite{Bartolo:2004if,Chen:2010xka} for reviews. A particularly significant and straightforward category is local-type non-Gaussianities, which typically arise when more than one light degrees-of-freedom are present during inflation, known as multi-field inflation. A notable prediction of multi-field inflation is $f_{\rm NL}\gtrsim1$; see e.g. Ref.~\cite{Alvarez:2014vva}. In comparison, the current best constraint from the latest \textit{Planck} satellite CMB analysis~\cite{Planck:2019kim} is 
$f_{\rm NL}= -0.9 \pm 5.1$. However, the primary CMB constraints are nearing their cosmic variance limit, as a significant portion of the available modes in temperature and polarization have already been measured.

For local non-Gaussianity, constraints can also be derived from the large-scale distribution of galaxies by analyzing the galaxy power spectrum in the linear regime. This approach leverages the fact that a nonzero $f_{\rm NL}$ introduces a scale-dependent galaxy bias~\cite{Dalal:2007cu}, which creates a distinct signal that is to a large extent not replicated by variations in other standard cosmological parameters or astrophysics. 

The effect on the galaxy bias can be calculated by noting first the Fourier-space relation between the primordial potential and matter overdensity
field $\delta_m(k, z) = \alpha(k, z)\Phi(k)$, where the form of the Poisson equation based operator $\alpha$ is given by 
\be
\alpha(k,z)=\frac{2k^2T(k)G(z)}{3\Omega_m H_0^2}
\ee
Here, $G(z)$ is the linear growth rate normalized such that
$G(z) = 1/(1 + z)$ during matter domination and $T(k)$ is
the transfer function normalized to 1 at low $k$.

In the presence of primordial non-Gaussianity, the number density of halos varies not only
with the large-scale matter overdensity modes but also
with the local small-scale power. This can be accounted
for in the derivation of the Lagrangian halo bias, by setting the non-Gaussian bias in~Eq.~\eqref{eq:nongawinf} to $b_{\rm NG}(k,z)=\beta_f/\alpha(k,z)$ where $\beta_f=2\partial \ln{n}_g/\partial \ln\sigma$. Here, ${n}_g$ is the galaxy number density and $\sigma$ is the local small scale matter power. We set $\beta_f\equiv2\delta_c(b_g-1)$~\citep{Barreira:2019qdl,2013PhRvD..88b3515S,Slosar:2008hx,Matarrese:2008nc,Afshordi:2008ru,2013MNRAS.435..934F} where $b_g$ is the galaxy bias, and take $\delta_c = 1.42$, as appropriate for the Sheth-Tormen halo mass function~\citep{2001MNRAS.323....1S}.

\begin{table}[t!]
    \begin{tabular}{|c|c|c|}
    \hline
         CMB & $68\%$ & $95\%$ \\
         \hline
         \multirow{2}{*}{\texttt{Commander}}& 
         $-287\lesssim f_{\rm NL}\lesssim 17$ & $-478\lesssim f_{\rm NL}\lesssim 206$ \\
         & $-29\lesssim A_{\rm CIP}\lesssim 367$ & $-242\lesssim A_{\rm CIP}\lesssim 568$ \\ 
         \hline
         \multirow{2}{*}{\texttt{SMICA}} & 
         $-220\lesssim f_{\rm NL}\lesssim 136$ & $-409\lesssim f_{\rm NL}\lesssim 335$ \\
         & $-147\lesssim A_{\rm CIP}\lesssim 281$ & $-384\lesssim A_{\rm CIP}\lesssim 509$ \\ 
         \hline
    \end{tabular}
    \caption{Constraints on $A_{\rm CIP}$ and $f_{\rm NL}$ from $\hat{C}_\ell^{\hat{v}g}$ using \texttt{Commander} and \texttt{SMICA}. We assume a flat prior on $b_v$ within $\in[0.5,1.1]$.}
    \label{tab:fnl_acip_b1}
\end{table}
\begin{table}[t!]
    \centering
    \begin{tabular}{|c|c|c|}
    \hline
         CMB & $68\%$ & $95\%$ \\
         \hline
         \multirow{2}{*}{\texttt{Commander}} & $-295.\lesssim f_{\rm NL}\lesssim 22$ & $-501\lesssim f_{\rm NL}\lesssim 223$ \\
         & $-29\lesssim A_{\rm CIP}\lesssim 367$ & $-242\lesssim A_{\rm CIP}\lesssim 567$ \\ 
         \hline
         \multirow{2}{*}{\texttt{SMICA}} & $-219\lesssim f_{\rm NL}\lesssim 120$ & $-434\lesssim f_{\rm NL}\lesssim 328$ \\
         & $-147\lesssim A_{\rm CIP}\lesssim 281$ & $-385\lesssim A_{\rm CIP}\lesssim 508$ \\ 
    \hline
    \end{tabular}
    \caption{Similar to Table~\ref{tab:fnl_acip_b1}. We assume a flat prior on $b_v$ within $b_v<4$. }
    \label{tab:fnl_acip_b2}\vspace*{-0.4cm}
\end{table}

\section{Posteriors for CIPs and PNG}

As above, our analysis is based on QML estimates for the full-sky reconstruction and galaxy density power spectra. We construct the posteriors over $f_{\rm NL}$ and $A_{\rm CIP}$ as follows. For a range of parameters $f_{\rm NL}$, $A_{\rm CIP}$ and velocity bias values $b_v$, we calculate the signal spectra $\{C_\ell^{vv}, C_\ell^{vg}, C_\ell^{gg}\}$ assuming standard \textit{Planck} $\Lambda$CDM as discussed in the text. The sampled parameter ranges are $-600<f_{\rm NL}<600$ (or $-600<A_{\rm CIP}<600$) and $0<b_v<6$. We use these spectra along with simulations of (uncorrelated) reconstruction noise, large-scale galaxy systematics and shot noise, to generate $10^4$ Gaussian maps of the remote dipole field and galaxy density using the \texttt{healpy} \texttt{synfast} function. We apply the unWISE galaxy masks to these maps and estimate the cross- and auto-correlation spectra of velocity and density for each simulation using Eq.~\eqref{eq:spec_est}. For a given multipole, we calculate the likelihood of the velocity and galaxy power-spectra by calculating the probability distribution of the estimated spectra values as a function of parameters $f_{\rm NL}$, $A_{\rm CIP}$ and $b_v$. We assume a uniform $b_v$ prior satisfying $0.5<b_v<1.1$ or $b_v<4.0$ and marginalize over $b_v$ to calculate 1-d marginalized posterior on parameters $f_{\rm NL}$, $A_{\rm CIP}$. 

\begin{figure*}[t!]
    \centering
    \includegraphics[width=0.96\textwidth]{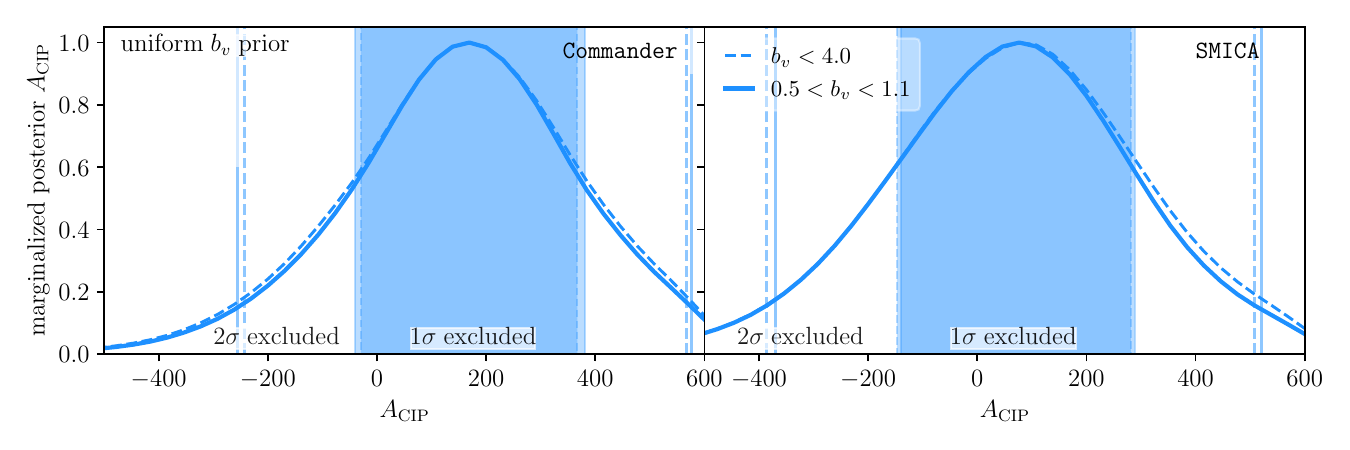}\vspace{-0.4cm}
    \includegraphics[width=0.96\textwidth]{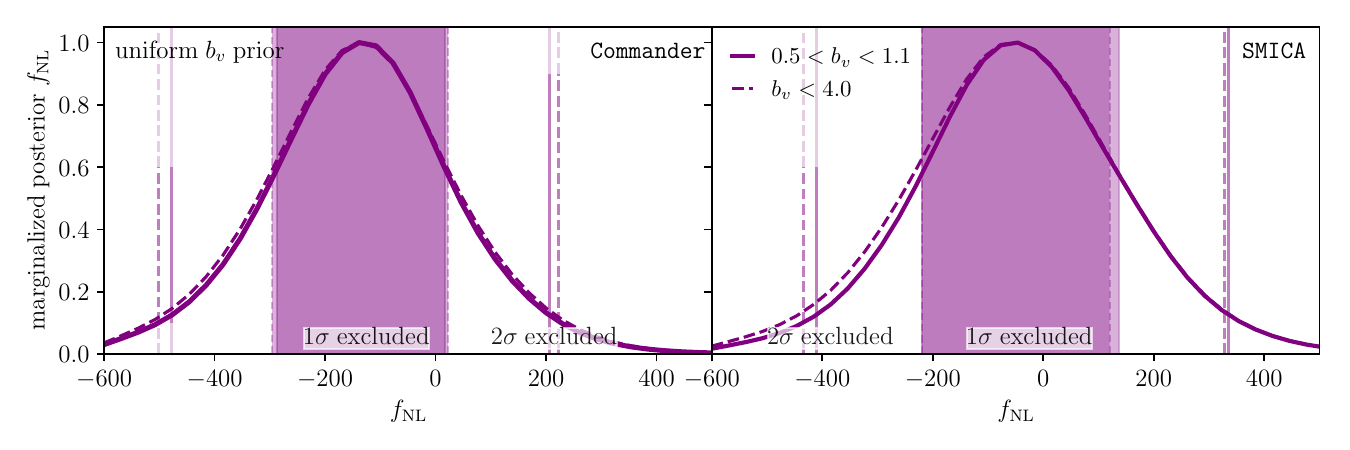}\vspace{-0.6cm}
    \caption{Constraints from \texttt{Commander} (left panel) and \texttt{SMICA} (right panel) for $A_{\rm CIP}$ in blue and $f_{\rm NL}$ in purple. We used two choices of priors for the optical depth, the solid line is over the range $0.5\leq b_v \leq 1.1$ and the dashed line is for $b_v < 4$. The vertical shaded regions (lines) enclose the 68\% (95\%) interval.}
    \label{fig:fnl_and_cip_constraints}
\end{figure*}

Our 1-d posterior results are shown in Fig.~\ref{fig:fnl_and_cip_constraints} for our two prior choices and for two CMB foreground cleaning methods, \texttt{SMICA} and \texttt{Commander}. Here we use only the velocity-galaxy cross-correlation spectra at multipoles ranging within $\ell\in[2,20]$. We find $\gtrsim\mathcal{O}(10)$ sensitivity in our 1- and 2-sigma confidence limits to the choice of foreground handling of the CMB. We find weak dependence of the results on the $b_v$ prior choice using \texttt{SMICA}, while with \texttt{Commander} our posterior distribution for large and negative $f_{\rm NL}$ values change by around a factor 2, $\sim1\sigma$ away from the maximum of the posterior. Note that these sensitivities are driven by the cross power spectrum between the reconstructed-velocity and unWISE galaxy density on large scales where reconstruction noise is expected to dominate over the signal. 

\end{document}